\documentclass[twocolumn]{aastex631}

\usepackage{color}

\shorttitle{BH NS mergers from wide systems}
\shortauthors{Michaely and Naoz}

\begin{document}

\title{Ultra wide black-hole - neutron star binaries \\ as a possible source for gravitational waves and short gamma ray
bursts}


\correspondingauthor{Erez Michaely}
\email{erezmichaely@gmail.com}

\author[0000-0002-9705-8596]{Erez Michaely}
\affiliation{Department of Physics and Astronomy, University of California,
Log Angeles, CA 90095, USA}
\affiliation{Mani L. Bhaumik Institute for Theoretical Physics, University
of California, Log Angeles, CA 90095, USA}

\author[0000-0002-9802-9279]{Smadar Naoz}
\affiliation{Department of Physics and Astronomy, University of California,
Log Angeles, CA 90095, USA}
\affiliation{Mani L. Bhaumik Institute for Theoretical Physics, University
of California, Log Angeles, CA 90095, USA}

\begin{abstract}
{ The third observing run of the LIGO/Virgo/KARGA collaboration reported a few neutron star–black hole (NSBH) merger events. While NSBH mergers have yet to receive extensive theoretical attention,  they may have a promising electromagnetic signature in the form of short gamma-ray bursts. Here we show that NSBH dynamical mergers can naturally form from ultra-wide binaries in the field. Flyby gravitational interactions with other neighbors in the galaxy in these ultra-wide systems may result in high eccentricity that drives the binary into a merger. We show that this process can result in a merger rate at the order of $\sim 10$~Gpc$^{-3}$~yr$^{-1}$ ($\sim 5$~Gpc$^{-3}$~yr$^{-1}$) for elliptical (spiral) galaxies. This channel predicts higher merger rate with higher velocity dispersion of the host-galaxy, delay time distribution which shallower than uniform but steeper that $1/t$, higher merger rate for lower BH to NS mass ratio. }


\end{abstract}


\section{Introduction} \label{sec:intro}

With the recent third catalog of observed gravitational waves signals
was published \citep{LSC2021}, a total of $90$ GW candidates
were detected, which allows a significant improvement of the inferred
merger rates for each of the three classes of binary systems, Binary
Black-hole (BBH), Binary neutron stars (BNS) and black-hole (BH) neutron
star (NS) binary (BHNS). The LIGO-VIRGO-KARGA collaboration \citep{LSC2021a} report that the for the BBH class the merger rate is $17.3-45{\rm Gpc^{-3}yr^{-1}}$,
for the BNS the inferred rate is $13-1900{\rm Gpc^{-3}yr^{-1}}$ and
for the BHNS the corresponding rate is $7.4-320{\rm Gpc^{-3}yr^{-1}}.$

{The formation of GW sources can roughly be divided into four groups:}

{\bf (1)  Isolated stellar binary:}  In this channel,  massive stellar binaries evolve together, in the absence of additional dynamical interaction \citep[e.g.,][]{Dominik2015,deMink2015,Belczynski2016,Eldridge2017,Giacobbo2018,Olejak2020}, or via stable mass transfer \citep[e.g.][]{GallegosGarcia2021}.  Fraction of the close massive binaries evolve through one or two common envelope process resulting in a possible formation of short period binary. Some fraction of these system are sufficiently close to merge by the emission of GW within Hubble time. A well known challenge to this avenue is the common envelope phase which may cause the binary to merge before the formation of the compact objects. However, a proposed avenue that may overcome this challenge may lay in the form of chemically
homogeneous stellar evolution which may avoid mass transfer  \citep[e.g.,][]{Marchant2016,deMink2016,Mandel2016}.  

{\bf (2) Dynamical mergers in dense environments.} Dense environments are efficient in producing many binary exotica due to the high stellar density. Galactic centers, 
or globular clusters and even low mass young clusters and open clusters are heavily populated with stellar and compact objects. In these environments, stars and compact objects, both singles and binaries experience strong gravitational interactions with individual stars or with higher multiplicity systems.
These interactions tend to
harden hard binaries, and possibly unbind soft binaries \citep{heggie1975}, whilst change their eccentricities. 
As a results some fraction of
these binaries may merge via GW emission in a Hubble time \citep[e.g.,][]{antonini2012,Samsing2014,Rodriguez2016,Rodriguez2018,Rodriguez2021,Fragione2018,Banerjee2018,Hamers2018,Leigh2018,Kremer2020,Ye2020,Stephan2019,Hoang2020,Wang2020,Rastello2021,Fragione2021}. Furthermore, it was suggested that AGN disks may assist with the merger of BH binaries \cite[e.g.,]{Bartos2017,Stone2017}.

{\bf (3) Secular evolution in hierarchical triple systems.} This channel considers  the long term evolution of triple body system via the Eccentric Kozai-Lidov (EKL) mechanism  \citep[e.g.,][see latter for review]{Lidov1962,Kozai1962,Naoz2016}.  

The EKL mechanism drive the inner binary to high eccentricity, which boost the GW emission power of the inner binary and lead to a merger. 
Triples are hosted either in the field of the host galaxy  \citep[e.g.,]{Antonini2016,Antonini2017,Silsbee2017,Liu2018,VignaGomez2021}
or in dense environments  \citep[e.g.,]{Britt2021,Miller2002a,Antonini2014,Kimpson2016,Samsing2018,Hamilton2019,Martinez2020}.
Moreover, at the center of galaxies, the third object can be the supermassive
BH \citep[e.g.,][]{antonini2012,Stephan2016,Stephan2019,Petrovich2017,Hoang2018,Fragione2019,Wang2020}.

{\bf (4) Mergers from ultra wide systems in the field.} Recently, it was shown that isolated ultra-wide systems (with semi-major axis (SMA)  $> 100\rm AU$),  either binaries or triples, in the field of the host galaxy, may be driven to extremely high eccentricities to allow strong interaction between the two binary objects \citep{Michaely2016,Michaely2019,Michaely2020,Michaely2020b,Michaely2021}.  For ultra-wide systems, the field is collisional environment due to consecutive frequent flyby interactions with random field stars. These interactions may excite the eccentricity (in the case of binaries) or outer eccentricity (in case of triples), with the result that mergers can occur via increased GW emission (binaries) or three-body instabilities (triples). 

Here we focus on this latter channel and study the BHNS formation channel. In particular we present a novel scenario that can lead to a BHNS merger via GW and a possibility to a EM counterpart.  

In the last years the majority of the literature focused on the BBH formation channels. 
However, recently a few studies have focused on identifying the possible origins of these BHNS mergers, \citep[e.g.,][see former study's introduction for an overview of the field]{Hoang2020,Ye2020,Mandel2021}.

BHNS mergers are uniquely interesting due to the possibility of an electromagnetic (EM) counterpart signal caused by the disruption of the NS from the BH tidal forces. This might result in a short-Gamma ray burst  \citep[sGRB, e.g.,][]{Berger2014} and other EM transients. sGRBs are intense, non-repeating and short flashes of gamma rays caused by the disruption of a NS due to tidal forces from its companion (either a NS or a BH, in this manuscript we focus on the BH-NS binaries). Unlike, long duration GRBs (longer than 2 seconds), which are associated with massive stars or core-collapse supernovae, that
are observed solely in star forming galaxies, sGRBs are observed in
both, star forming galaxies and elliptical galaxies. Overall about $20\%$
of sGRBs are found in early-type (elliptical and $S0$ galaxies) \citep[e.g.,][]{Leibler2010,Berger2014}.
Here we show that our novel channel naturally predicts $\sim 1\% - 50\%$ of all BH-NS mergers to be accompanied with sGRB transient.

This paper is organized as follows. In section \ref{sec:Physical Picture} we preset the physical picture, derive the governing rate equations and calculate the merger rates. We discuss the results and summarize the manuscript in section \ref{sec:Discussion}.

\section{\label{sec:Physical Picture}Physical Picture}

\subsection{The Collisional Nature of Ultra-Wide Binaries in the Field}

This section describes the merger scenario and calculates the merger rate for a Milky-Way (MW)-like galaxy and an elliptical galaxy.  Although low in stellar density, the galactic field is considered collisional for ultra-wide systems. These wide systems interact relatively frequently with random flyby stars in the field because of their large interaction cross-section proportional to their SMA. Hence, an ultra-wide system encounters many random interactions with flyby field stars. 

These interactions are typically impulsive by nature, i.e., the interaction timescale $t_{{\rm int}}\equiv b/v_{{\rm enc}}$
(where $b$ denotes the closest approach to the binary's center of
mass, and $v_{{\rm enc}}$ denotes the relative velocity between the
random flyby star and the binary) is much shorter than the orbital
period time, $P$. Moreover, these interactions change the binary
orbit characteristics, namely the orbital energy (decrease/increase
the SMA), but more significantly, torque the system and change the
binary eccentricity $e$ \citep[e.g.,][]{Lightman1977,Merritt2013}. After
many impulsive interactions the wide binary might be driven into a
sufficiently high eccentricity, and as a result to sufficiently small
pericenter distance, $q$,  which allows the binary to merge via GW
emission within Hubble time (or less). Once a BH-NS mergers via GW we might expect an EM transient associated with the mergers, namely a sGRB.

\subsection{Relevant Timescales  \label{subsec:Timescales}}

As mentioned above in this scenario, the impulse interaction timescale $t_{{\rm int}}\equiv b/v_{{\rm enc}}$ needed to be shorter than the binary orbital period $P$. Specifically we require that  $t_{{\rm int}}= \alpha P$, with $\alpha \ll 1$.  This gives upper bound to the closest approach distance, $b_{\rm{impulse}}=\alpha Pv_{\rm enc}$ and hence limits the average time between encounters. For example, consider a binary with total mass of $10M_{\odot}+1.5M_{\odot}=11.5M_{\odot}$, SMA of $a\sim10^{4}{\rm AU}$ which translates to an orbital period of $P\approx3.2\times10^{5}{\rm yr}$ and velocity encounter of $v_{{\rm enc}}=50{\rm kms^{-1}}$ we can restrict $b_{\rm{impulse}}$ for $\alpha = 0.1$. Thus, we get $b_{\rm impulse}= \alpha t_{{\rm int}}\times v_{{\rm enc}} \sim 3.3\times10^{5}{\rm AU}$.

Additionally, we consider the average time between two flybys of the binary system and a random stellar perturber, $t_{{\rm enc}}=1/f=\left(n_{*}\sigma v_{{\rm enc}}\right)^{-1}$ where $n_{*}$ denotes the stellar number density, $\sigma=\pi b^{2}$
is the geometric cross-section of the binary and the stellar fly-by, and $v_{{\rm enc}}$ is the relative velocity of the fly-by and the
binary center of mass. 

Lastly, we consider the GW timescale \citep[e.g.,][]{Pet64}
\begin{equation}
t_{{\rm merger}}\approx\frac{a^{4}}{\beta'}\times\left(1-e^{2}\right)^{7/2}\label{eq:t_merger}
\end{equation}
where 
\[
\beta'=\frac{85}{3}\frac{G^{3}m_{{\rm BH}}m_{{\rm NS}}\left(m_{{\rm BH}}+m_{{\rm NS}}\right)}{c^{5}}
\]
$G$ is Newton's constant and $c$ is the speed of light. Thus, we focus on the merger of a wide binary with semi-major axis, $a>10^{2}{\rm AU}$. 



\subsection{\label{subsec:Analytic-Description}Analytic Description of the BH-NS merger}

Consider an ensemble of isotropic wide binaries, each consisting of a BH with mass $m_{{\rm BH}}=10$ and a NS with mass $m_{{\rm NS}}=1.5M_{\odot}$. We assume all binaries to have the same SMA, and a thermal distribution of orbital eccentricities, $f(e)de=2ede$. We derive the merger probability from this ensemble and find its dependence on the SMA of the binaries, $a$ for a given environment, i.e. stellar density $n_{*}$ and velocity dispersion.

We start by calculating the fraction of system that merge from this ensemble given thermal distribution of eccentricity using Equation (\ref{eq:t_merger}).

Given the
merger time $t_{{\rm merger}}=T$ and the SMA of the binary $a$ we define a 
critical eccentricity $e_{c}$
\begin{equation}
e_{c}=\left[1-\left(\frac{\beta'T}{a^{4}}\right)^{2/7}\right]^{1/2},
\end{equation}
for which all systems with eccentricities equal or greater than $e_{c}$ merge
within time $T$. Below we will adopt $T$ as the average encounter time. 
Thus, given a thermal distribution of eccentricities
we find the fraction of systems that merger within $T$ and lost from
the ensemble to be:
\begin{equation}
F_{q}=\int_{e_{c}}^{1}2ede=1-e_{c}^{2}=\left(\frac{\beta'T}{a^{4}}\right)^{2/7} \ .\label{eq:F_q}
\end{equation}
All other binaries outside the loss-cone are susceptible for interaction
with a random flyby and can be perturbed to change their angular momentum
and replenish the loss cone.

After a single random flyby interaction the relative velocity vector of the binary changes. The new direction of the velocity vector maps a cone around the original relative velocity vector. The average \emph{fraction} of the phase-space region into which the binaries in the ensemble are perturbed after a single fly-by is termed the smear cone. The smear cone is computed by the ratio of the size of the cone with all possible directions \citep[e.g.,][]{kaib2014,Michaely2016,Michaely2019}.  In order to calculate the smear cone one should first find the size of which , and defined by the following:
\begin{equation}
\theta=\frac{\left\langle \Delta v\right\rangle }{v_{k}} \ ,\label{eq:theta-1}
\end{equation}
 where $v_{k}$ is the Keplerian velocity of the binary and $\left\langle \Delta v\right\rangle $
is the average change of $v_{k}$ from a single perturbation. The
value of $v_{k}$ is calculated at the average separation of a Keplerian
orbit is $\left\langle r\right\rangle =a\left(1+1/2e^{2}\right)$
and we approximate $e\rightarrow1$, yielding
\[
v_{k}\sim  \left(\frac{GM}{3a}\right)^{1/2} \ ,
\]
where $M\equiv\left(m_{{\rm BH}}+m_{{\rm NS}}\right)$ is the total
mass of the binary.

The average change of the Keplerian velocity, $\left\langle \Delta v\right\rangle $
is given by \citep[e.g.,][]{Hills1981} 
\begin{equation}
\left\langle \Delta v\right\rangle \simeq\frac{3Gam_{p}}{v_{{\rm enc}}b^{2}},\label{eq:Delta_V}
\end{equation}
where $v_{{\rm enc}}$ is the velocity of the fly-by star with respect to the binary center of mass, $m_{p}$ is the perturber mass. 
Therefore, the square of the angular size of the smear cone cause by the impulse
of the fly-by on the binary is (using Equations (\ref{eq:theta-1}) -  (\ref{eq:Delta_V})) 

\begin{equation}
\theta^{2}=\frac{9G^{2}a^{2}m_{p}^{2}}{\left(v_{{\rm enc}}b^{2}\right)^{2}}\frac{3a}{GM}=\frac{27Ga^{3}m_{p}^{2}}{M\left(v_{{\rm enc}}b^{2}\right)^{2}} \ ,\label{eq:theta_calc}
\end{equation}
and for $\theta\ll1$ we get the average size of the cone post interaction. The smear-cone is the fraction of that size with all possible directions, i.e.  over the $4\pi$ sphere to be after a single passage
of the perturber \citep[e.g.,][]{kaib2014,Michaely2016,Michaely2019}
\begin{equation}
F_{s}=\frac{\pi\theta^{2}}{4\pi}=\frac{27}{4}\left(\frac{m_{p}}{M}\right)^{2}\left(\frac{GM}{av_{{\rm enc}}^{2}}\right)\left(\frac{a}{b}\right)^{4} \ .\label{eq:F_s}
\end{equation}
For a given binary, the smear cone size in Equation (\ref{eq:F_s}) depends on
the perturber and the environment, specifically, its mass, $m_{p}$, closest
approach, $b$ and the encounter velocity, $v_{{\rm enc}}.$

The ratio between the smear and loss cones, is the fraction of the
loss cone filled after a single flyby. Namely, for cases where $F_{s}\ge F_{q}$
the entire ensemble regain its thermal eccentricity distribution.
\begin{equation}
\frac{F_{s}}{F_{q}}=\frac{27}{4}\left(\frac{m_{p}}{M}\right)^{2}\left(\frac{GM}{av_{{\rm enc}}^{2}}\right)\left(\frac{a}{b}\right)^{4}\left(\frac{a^{4}}{\beta'T}\right)^{2/7} \ .\label{eq:Fs/Fq}
\end{equation}
The are two conditions for the loss cone to be continuously full is
(1) that the kick is sufficiently strong to fill the loss cone, namely, $F_{s}\ge F_{q}$, and (2) the replenishment is as fast as the orbits are lost from the ensemble. The replenishment rate is determined by the interaction rate which can easily by calculate by
\begin{equation}
f\equiv\frac{1}{t_{{\rm enc}}}=n_{*}\sigma\left(b\right) v_{{\rm enc}}. \label{f_t_enc}
\end{equation}

\begin{figure}
\includegraphics[width=1\columnwidth]{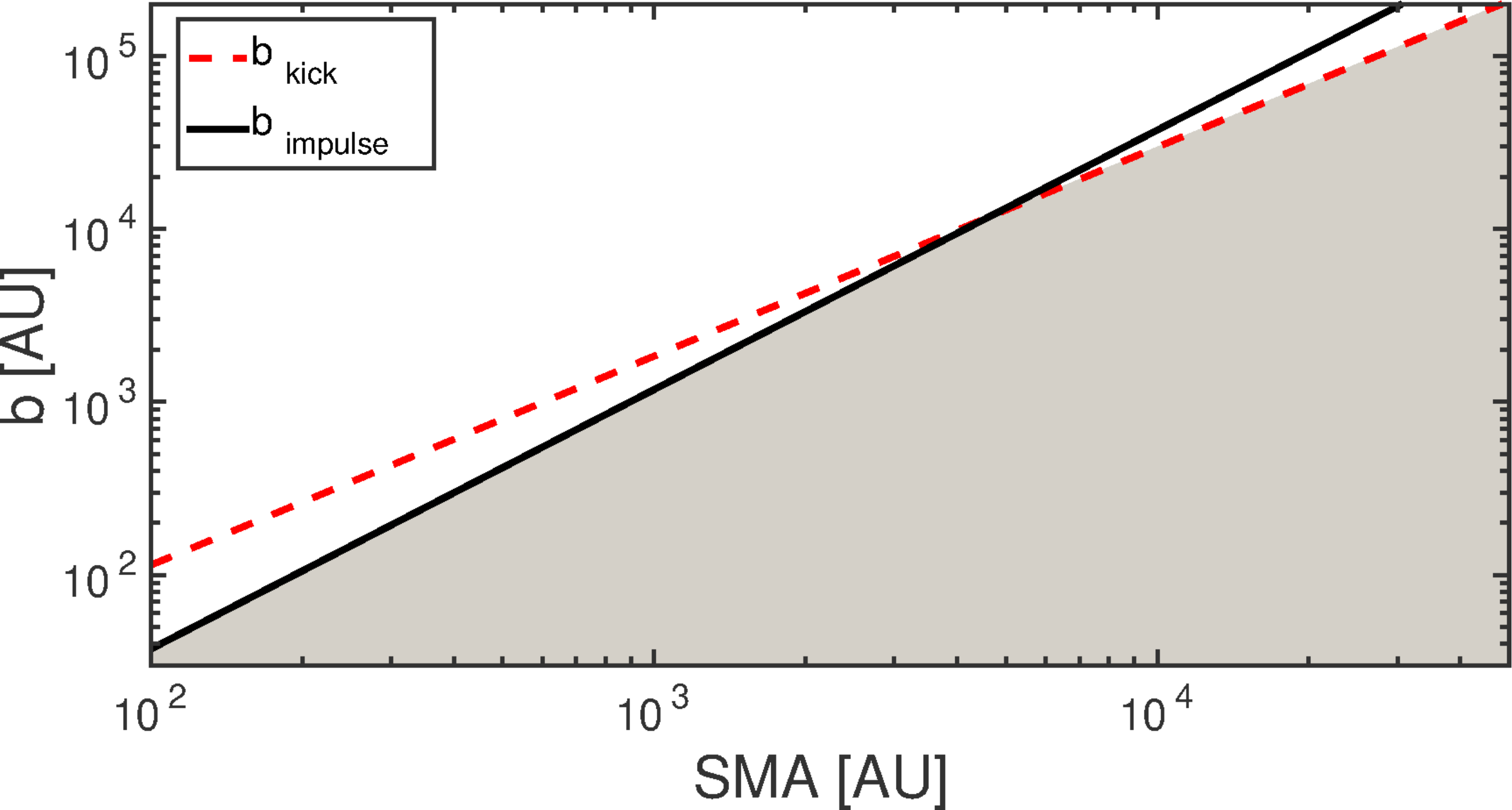}
\caption{\label{fig:b_vs_a}The two distance scales $b_{\rm kick}$ (dashed red) and $b_{\rm impulse}$ (solid black). Each fly-by with $b<b_{\rm kick}$ will fill the loss cone and each one with $b<b_{\rm impulse}$ will interact impulsively with the wide BHNS binary. We are interested in the grey area which insures both conditions. Calculated for binary with $m_{{\rm BH}}=10M_{\odot}$ and $m_{{\rm NS}}=1.5$ with $v_{{\rm enc}}=60{\rm kms^{-1}}$.
The stellar density number is $n_{*}=0.1{\rm pc^{-3}}$. }
\end{figure}

The first condition, the demand that the flyby kick is sufficiently strong, sets a limit on the closest approach, 
\begin{equation}
b_{\rm kick}^2 \le v_{\rm enc}^{-1}\left(\frac{27G}{4} a^{29/7} \frac{m_p^2}{M} \left( \frac{n_*\pi}{\beta'} \right)^{2/7}\right)^{7/12}. 
\label{eq:b_kick}
\end{equation}
We set the close approach to be the minimum of the two distance scales, the first $b_{\rm impulse}$ is required because we focus only on the impulsive regime, the second $b_{\rm kick}$ is required to ensure each fly-by is sufficiently strong to fill the loss cone. Specifically,
\begin{equation}
b=min\left(b_{\rm kick},b_{\rm impulse}\right),
\label{eq:b_min}
\end{equation} see Figure \ref{fig:b_vs_a} for visualization.

Next, we can calculate the average encounter time, $t_{\rm enc}$ from Equation (\ref {f_t_enc}). We chose the merger time to the average time between encounters, which is a function of the closest approach, $T\equiv t_{\rm enc}\left(b\right)$. Note that the merger time $T$ is dependent on the closest approach $b$, which is determined from Equation (\ref{eq:b_min}).

Now we are set to find the critical SMA from which the loss cone is continuously full, under the impulse approximation, as a function of the local environment characteristics. Thus, for $F_{s}=F_{q}$ and $1/t_{{\rm enc}}=n_{*}\sigma v_{{\rm enc}}$, we find: 

\begin{equation}
    a_{{\rm crit}}=\left[\frac{4}{27}\frac{M\beta'^{2/7}T\left(b \right)^{-12/7}}{Gm_{p}^{2}n_{*}^{2}\pi^{2}}\right]^{7/29}.
\end{equation}

Another way to express the critical SMA is without the specific explicit function of the geometric cross section, $\sigma$

\begin{equation}
a_{{\rm crit}}= \left(\frac{GM}{n_{*}^{2}\sigma(b)^2v_{\rm enc}^{2}} \right)^{1/3}
\label{eq:a_crit2}
\ .
\end{equation} 
Binaries with SMA than equal to the $a_{{\rm crit}}$ have the highest probability to merger via GW emmision.  Binaries with SMA smaller than $a_{{\rm crit}}$ are harder to perturb, therefore the probability rate decreases. Binaries with SMA larger than $a_{{\rm crit}}$ also descreas in probability due to the smaller loss cone, additionally, these binaries are more likely to become unbind during the interaction, e.g. \citet{Michaely2019}]. This interplay leads to a sharp transition in the merger probability, as highlighted in Figure \ref{fig:The-merger-probability}, see below. 

In other words, the full loss cone regime is for all systems with $a>a_{{\rm crit}}$. The loss rate of systems from the full loss-cone, ($\dot{L}_{{\rm full}}$), is given by the rate
they leave the loss cone
\begin{equation}
\dot{L}_{{\rm full}}=\frac{F_{q}}{P} \ .\label{eq:Loss_rate-1}
\end{equation}
On the other hand, tighter binaries are less susceptible for change due to a fly-by, this is evident from equation (\ref{eq:F_s}). Therefore for $a<a_{{\rm crit}}$ we expect that the loss cone will not be full all the time, in this ``empty loss cone'' the loss rate depends on the rate of orbits being kicked into the loss cone is just the interaction rate, $f$:
\begin{equation}
\dot{L}_{{\rm empty}}=F_{q}f=\frac{F_{q}}{t_{{\rm enc}}}
\end{equation}
From the critical SMA we can calculate the merger probability for each regime, for $a<a_{{\rm crit}}$ (empty loss cone) and for $a>a_{{\rm crit}}$
(full loss cone). We note here that by considering the critical SMA we implicitly include the dependency on the impact parameter, $b$.

In what follows we calculate the time dependent loss probability of the two regimes, the empty and full cones.
We note that $F_q$ is the fraction of systems that are lost from the ensemble and therefore $\left(1-F_{q}\right)$ represents the fraction of binaries
that survive as wide binaries at the relevant timescale,  after a single flyby with the relevant timescale. Therefore, $\left(1-F_{q}\right)$ is a monotonically decreasing function of time. Hence, the surviving fraction of binaries after time $t$ and the relevant timescale for the empty loss cone $f$ is $\left(1-F_{q}\right)^{tf}$ \citep[e.g.,][]{kaib2014,Michaely2016,Michaely2019}. Therefore the probability for a wide binary merger, which complements this expression to unity is
\begin{equation}
L_{a<a_{{\rm crit}}}=1-\left(1-F_{q}\right)^{tf}\label{eq:empty_Prob}
\end{equation}
where $t$ is the time since birth of the binary. As one can expect
the probability only depends on the size of the loss cone and the
rate of interactions. For the limit of $tfF_{q}\ll1$
we can expand Equation (\ref{eq:empty_Prob}) and take the leading
term, to find the loss probability to be approximated by 
\begin{equation}
L_{a<a_{{\rm crit}}}=tf(b)F_{q}.\label{eq:empty_prop_approx}
\end{equation}

For the full loss cone regime the limiting factor is not the value of $f$, but rather the orbital period $P$. Therefore, the full expression for the loss probability for $a>a_{crit}$ is
\begin{equation}
L_{a>a_{{\rm crit}}}=1-\left(1-F_{q}\right)^{t/P}.\label{eq:full_prob}
\end{equation}
 For the limit $F_{q}\cdot t/P\ll1$ we can approximate the probability by 
\begin{equation}
L_{a>a_{{\rm crit}}}=tF_{q}~\frac{1}{P(a)}.\label{eq:full_prob_apx_express}
\end{equation}

\begin{figure}
\includegraphics[width=1\columnwidth]{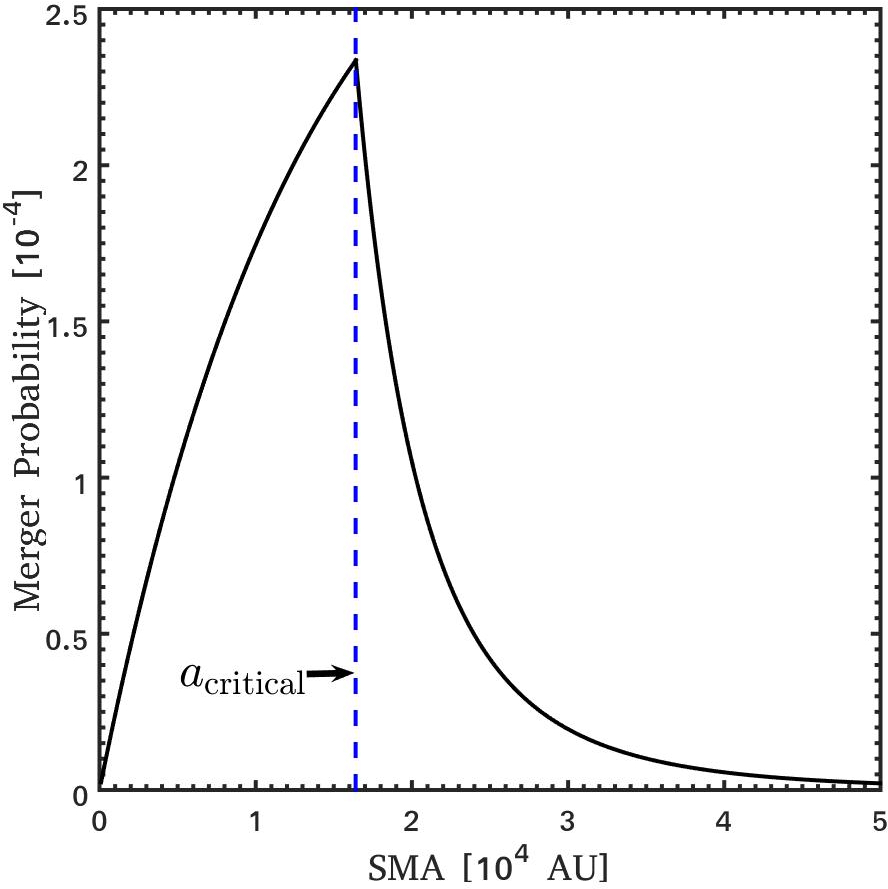}
\caption{\label{fig:The-merger-probability}The merger probability of BHNS
with $m_{{\rm BH}}=10M_{\odot}$ and $m_{{\rm NS}}=1.5$ with fly-by
mass $m_{p}=0.6M_{\odot}$ and $v_{{\rm enc}}=50{\rm kms^{-1}}$.
The stellar density number is $n_{*}=0.1{\rm pc^{-3}}$. The probability
it calculated after $t=10{\rm Gyr}$ since the BHNS was formed. The
peak probability is achieved at $a=a_{{\rm crit}}$. For binaries
with tighter orbits the merger probability drops because it is harder
to change their orbital parameters. For binaries with $a>a_{{\rm crit}}$
the probability decreases because the loss cone dependency with $a$,
see text.}
\end{figure}

Until this point we neglect the ''ionization' of wide binaries
in the field. This process reduce the available number of wide binaries. To account for this we use the half-life time approximate relation given by \citep{Bahcall1985}, the half-life time of a wide binary
is 
\begin{equation}
t_{1/2}=0.00233\frac{v_{{\rm enc}}}{Gm_{p}n_{*}a}.\label{eq:t_half_life}
\end{equation}
Correcting for this we get for Equation (\ref{eq:empty_prop_approx})
and Equation (\ref{eq:full_prob_apx_express}) the following approximations:
\begin{equation}
L_{a<a_{{\rm crit}}}=\tau f(b)F_q\left(1-e^{-t/\tau}\right)\label{eq:ionization empty}
\end{equation}
and 
\begin{equation}
L_{a>a_{{\rm crit}}}=\tau \frac{F_q}{P(a)}\left(1-e^{-t/\tau}\right),\label{eq:ionization full}
\end{equation}
where $\tau=t_{1/2}/\ln2$ is the mean lifetime of the binary.  For a more detailed derivation see equations (11) and (12) in \citet{Michaely2019} and in the Appendix.

\subsection{\label{Sec:Volumetric rates}Volumetric merger rates}

In this subsection we calculate the GW signal rate of the local universe
from wide BH-NS binaries. In order to do that we need (1) to approximate
the fraction of wide BH-NS binaries out of the stellar population.
(2) To model the stellar density of the host galaxy.

First we estimate the fraction of wide BH-NS systems from the stellar
population. The initial binary is comprised out of two stars with
masses $M_{1}$ and $M_{2}$.  As a prove of concept  we assume that all stars with zero age
main sequence (ZAMS) masses greater than $20M_{\odot}$ turn into
BHs without any natal-kicks \citep[and others]{Belczynski2016,Mandel2016a}.  We emphasize that this is a simplifying assumption, it is currently unknown the mapping of ZAMS, metallicity and the stellar spin to the remnant BH mass.
Additionally, we assume that all stars with ZAMS between $8-10M_{\odot}$
turn into NS via electron capture Supernova (SN) and born without natal kicks
\citep{nomoto1982a,nomoto1984}. For Kroupa (Salpeter) initial
mass function (IMF) we find that $f_{{\rm BH}}\approx2\times10^{-3}\ \left(5\times10^{-4}\right)$
\citep{Kroupa2001,salpeter1955}. The companion mass
is calculated given a mass ratio distribution, $Q\equiv M_{2}/M_{1}$
with $Q\propto M_{{\rm 1}}^{-2}$ \citep{Moe2016} with the boundaries
$Q_{{\rm outer}}\in\left(0.3,1\right)$, this translate to $f_{{\rm NS}}\approx0.13\ \left(0.12\right)$.
We assume that the binary fraction $f_{{\rm binary}}=1$ \citep{Duchene2013}.
Next we estimate the fraction of binaries with SMA greater than $a>1000{\rm AU}$
to be $f_{{\rm wide}}\approx0.2$ \citep{Michaely2019}
\begin{equation}
f_{{\rm BH-NS}}=f_{{\rm BH}}\times f_{{\rm NS}}\times f_{{\rm binary}}\times f_{{\rm wide}}\approx5.2\times10^{-5}
\end{equation}
and for Salpeter IMF we get $f_{{\rm BH-NS}}\approx1.2\times10^{-5}$.

Next, we model the stellar density function for a Milky-Way like spiral
galaxy and for an elliptical galaxy. For the spiral case we follow
the model used in \citep{Michaely2016}. The number of stars in region
$dr$ at a distance $r$ from the center is 
\begin{equation}
dN(r)=n_{*}\left(r\right)~2\pi~r ~ h~ dr
\end{equation}
 where the scale height is $h$. We continue with modeling the Galactic
stellar density in the Galactic disk as follows 
\begin{equation}
n_{*}\left(r\right)=n_{0}e^{-\left(r-r_{\odot}\right)/R_{l}} \ ,\label{eq:MW_galaxy}
\end{equation}
 where $n_{0}=0.1{\rm pc}^{-3}$ is the stellar density in the solar
neighborhood, $R_{l}=2.6{\rm kpc}$ \citep{Juric2008} is the galactic
length scale and $r_{\odot}=8{\rm kpc}$ is the distance of the Sun
from the Galactic center. The mass of the perturber is taken to be
$0.6M_{\odot}$ which is the average mass of a star in the galaxy.
The encounter velocity is set to be the velocity dispersion of the
flat rotation curve of the galaxy, namely $\hat{\sigma}=50{\rm kms^{-1}}$.

For the elliptical galaxies, we use the density profile function from \citep{hernquist1990}, given the average stellar mass of
$0.6M_{\odot}$ we get
\begin{equation}
\tilde{n}_{*}\left(r\right)=\frac{M_{{\rm galaxy}}}{2\pi r}\frac{r_{*}}{\left(r+r_{*}\right)^{3}}\label{eq:elliptical}
\end{equation}
where $r_{*}=1{\rm kpc}$ is the scale length of the galaxy, $M_{{\rm galaxy}}=10^{11}M_{\odot}$
is the total stellar (not including dark matter) mass of the galaxy, we focus on $r=0.5-30\rm{kpc}$.
The elliptical galaxy velocity dispersion is taken to be $\hat{\sigma}=160{\rm kms^{-1}}.$

The merger rate is calculated by integrating the merger probability
for all SMA, with log-uniform distribution, $f_a$, and the entire host galaxy
\begin{equation}
\label{eq:Gamma_MW}
\Gamma_{{\rm MW}}=\int_{0.5{\rm kpc}}^{15{\rm kpc}}\int_{100{\rm AU}}^{5\cdot10^4{\rm AU}} \frac{L\left(a,r\right)}{T_{\rm int}} f_af_{{\rm BH-NS}}dadr \ 
\end{equation}
where $T_{\rm int}$ is the integration time, for $T_{\rm int}=10\rm{Gyr}$ the average merger rate, $\Gamma$ is for the lifetime of the host galaxy.  The integral limits are taken to be the relevant size of the MW galaxy and the all the binaries SMA we consider.
The function $L\left(a,r\right)$ is taken from eq. (\ref{eq:ionization empty})
and (\ref{eq:ionization full}). Thus, for the MW-like galaxy we get 
\begin{equation}
\Gamma_{{\rm MW}}\approx0.06{\rm Myr}^{-1} \ ,
\end{equation}
where, we use \citet{Belczynski2016} in order to calculate the merger rate, $R$
per ${\rm Gpc^{3}}$ by using the following equation 
\begin{equation}
{R_{{\rm MW}}}=10^{3}\rho_{{\rm gal}}\times\Gamma_{{\rm MW}}\approx0.66{\rm Gpc^{-3}yr^{-1}}
\end{equation}
while $\rho_{{\rm gal}}$ is local density of the MW-like galaxies
with the value of $\rho_{{\rm gal}}=0.0116{\rm Mpc^{-3}}$(e.g. \citet{Kopparapu2008})
and $\Gamma_{{\rm MW}}$ is given in the units of ${\rm Myr^{-1}.}$

For the elliptical case we compute:
\begin{equation}
\label{eq:Gamma_ellipti}
\Gamma_{{\rm elliptical}}= \int_{0.5{\rm kpc}}^{30{\rm kpc}}\int_{100{\rm AU}}^{5\cdot10^4{\rm AU}}\frac{L\left(a,r\right)}{T_{\rm int}} f_af_{{\rm BHNS}}dadr
\end{equation}
and get $\Gamma_{{\rm elliptical}}\approx0.39{\rm Myr^{-1}}$. 

The number density of elliptical galaxies in the local universe elliptical
is crudely approximated by $\sim0.1{\rm Mpc^{-3}}$ \citep{Samsing2014} and get
\begin{equation}
{R_{{\rm elliptical}}}\approx39{\rm Gpc^{-3}yr^{-1}.}
\end{equation}
Adding the contributions of each type of galaxy, we get 
\begin{equation}
    R_{\rm tot}=R_{\rm elliptical} + R_{\rm MW} \approx 40 \rm Gpc^{-3}yr^{-1}.
\end{equation}

\section{\label{sec:Discussion}Discussion and summary}

\subsection{Caveats and assumptions}

The mathematical model we present here is simplistic and based on
several assumptions. 

\noindent
\textbf{BH masses:} As mentioned in Sec. \ref{Sec:Volumetric rates}, we make simplifying assumptions on the remnant mass of the BH. The mapping of initial mass of the star, metallicity and spin to final BH mass is not well understood. The stellar evolution physics, e.g. wind mass loss, nuclear physics and supernova physics \citep{Baes2007,Zhu2010,Fryer2012}. Hence we present our results with the prefactor of $f_{\rm BH}=2\times10^{-3}$. Different assumption will lead to different fraction of BH from the initial populations.

\noindent
\textbf{BH Natal kicks:} The calculated merger rates depends strongly
on the binary fraction estimation. This estimation is stringently depended
on the natal kicks upon formation of BHs. These natal kicks are poorly
constrained \citep[e.g.,][]{Repetto2012,Repetto2017}. However,
there are evidence that BH are formed by failed SN \citep[e.g.,][]{ertl2015}. In the failed SN scenario large amount of fallback is
accreted on the newly formed compact object and suppresses any natal
kicks. Moreover, many theoretical studies of GW merger rate assume
not natal-kicks for BH \citep[e.g.,][]{belczynski2007,belczynski2008,Belczynski2016}. 

\noindent
\textbf{NS Natal kicks:} From the observations of pulsar it is well
known that NS are born with natal kicks. However, in the narrow range
of masses between $8-10M_{\odot}$, stars are believed to undergo
e-electron capture SN. In this work we limit ourselves only to this
narrow range of stars and assume no natal kicks for the newly born
NS.

\noindent
\textbf{SMA and eccentricity distributions} The underlying assumption
of the results presented here are the distributions of the SMA and
eccentricities of the wide systems. We use Opik law, namely, log-uniform
distribution for the SMA of the systems, and use thermal eccentricity
distribution for the eccentricity of the binaries.
\subsection{Model Predictions and Signatures}
In this study we explore a novel scenario that may lead to a merger between BH and a NS or a NS disruption. In what follows we present some of the model features.

\subsubsection{Velocity dispersion}  A unique signature of the model presented here is the dependence on the host galaxy velocity dispersion. The velocity dispersion sets the rate of encounters and the timescale of binary ionization. Following the model presented in the previous section we can calculate the scaling relation between the merger rate and the hot galaxy velocity dispersion. We distinguish between the regime where $b=b_{\rm {kick}}$ and the regime where $b=b_{\rm {impulsive}}$. For simplicity we start with the non-ionized case Equation (\ref{eq:empty_prop_approx}) and Equation (\ref{eq:full_prob_apx_express}). The scaling relation is calculated by considering each term dependency on $v_{\rm enc}$. The frequency $f\propto v_{\rm enc}$ from equation (\ref{f_t_enc}). The loss cone, $F_q\propto T^{2/7}a^{-8/7}$, noting that $a\propto v_{\rm enc}^{-2/3}$ from equation (\ref{eq:a_crit2}) we get $F_q\propto v_{\rm enc}^{10/21}$.  Therefore we get:
\begin{equation}
    L_{\rm impulse,empty } = tfF_q \propto v_{\rm enc}^{31/21}
    \label{eq:L_scale_sigma_impulse}
\end{equation}
The integrated rate, $\Gamma$, is the integral of (\ref{eq:L_scale_sigma_impulse}) over all SMA, $\int Lf_ada$.Hence, the scaling relation is

\begin{equation}
\label{gamma_scale_imp_not}
    \Gamma \propto L\frac{1}{a_{\rm crit}} a_{\rm crit}\propto v_{\rm enc}^{31/21}
\end{equation}

\begin{figure}
\includegraphics[width=1.0\columnwidth]{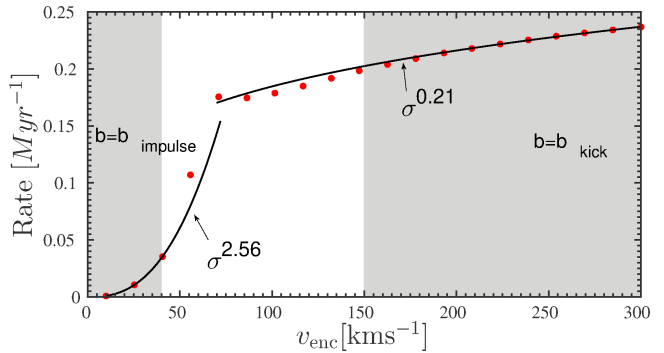}\caption{\label{Rate_Sigma} Average merger rate for a MW galaxy-like as a function of unrealistic velocity dispersion. Rate is calculated after $10^{10}\rm yrs$ from star formation accounting for binary ionization. Binaries consist of  $m_{\rm {NS}} =1.5M_{\odot}$ and $m_{\rm{BH}}=10M_{\odot}$. For $v_{\rm enc}\lesssim 40 \quad (\gtrsim 150) \quad \rm{kms^{-1}} $, the closest approach $b=b_{\rm impulse} \quad (b=b_{\rm kick})$ for all SMA (grey area), see text. The white area is a mix of both regimes. The rate for an elliptical galaxy shows a similar behavior and is omitted here to avoid clutter.}
\end{figure}

Including the ionization process, the merger probability changes to Eq. (\ref{eq:ionization empty}), where 
\begin{equation}
    \tau \propto \frac{v_{\rm enc}}{a} \ .
\end{equation}
For the case of $-t/\tau \rightarrow 0$ we get $\tau\propto v_{\rm enc}^0$ for the other extreme, $-t/\tau \rightarrow \infty$ we get $\tau \propto v_{\rm enc}^{5/3}$. 
For the regime where $b=b_{\rm kick}$, the calculation is similar but with different scaling. Specifically,
\begin{equation}
   L_{\rm kick,empty } = tfF_q \propto v_{\rm enc}^{0}.
    \label{eq:L_scale_sigma_kick}
\end{equation} 
As a result the $v_{\rm enc}$ dependency disappears
in the non-ionization case. Once we include ionization we get the following:
\begin{equation}
   \Gamma \propto \int \tau Lf_ada (1-e^{-t/\tau})\propto v_{\rm enc}(1-e^{-t/\tau}),
\end{equation}
where 
\begin{equation}
    \tau \propto v_{\rm enc} \ .   
\end{equation}
In Figure \ref{Rate_Sigma} we depict 
the merger rate scaling relation
as a function of the velocity encounter for a range of velocity dispersions. One can see the crossover in scaling around $v_{\rm enc} \approx 70~ \rm{km~s^{-1}}$ between the impulsive and the kick regimes.

Currently we cannot identify BHNS mergers for a specific galaxy, however, it is possible to assign a host galaxy to a sGRB coincident with GW detection.

\subsubsection{Delay time distribution}

\begin{figure}
\includegraphics[width=1\columnwidth]{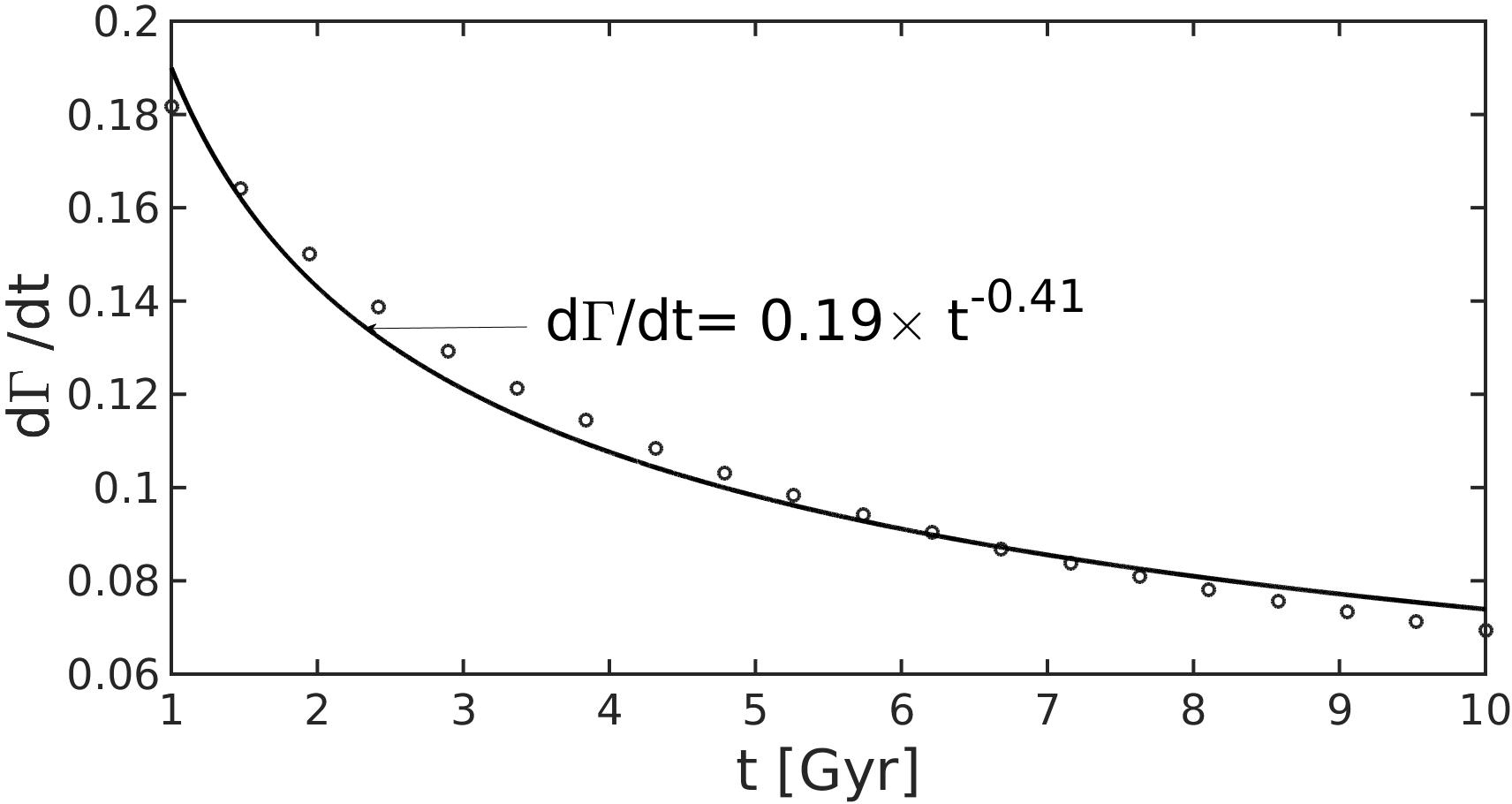}

\caption{\label{DTD} The Delay time distribution (DTD) for a MW-like galaxy with velocity dispersion of $\sigma=40\rm{kms^{-1}}$. Blue circles are the model calculations. Black solid line is the best power law fit. Note that this fit is specific for MW-galaxy adopted here, see text for more details.  The GW merger span for the entire lifetime of the galaxy, which suggests these mergers occur in early and late type of galaxies.}

\end{figure}

The BHNS delay time distribution (DTD), $d \Gamma /dt$, is the binary merger rate as a function of time since star formation. In this case we ignore the stellar evolution time, which is much quicker ($\sim 10^6 \rm yr$) than galaxy lifetime timescale ($\sim 10^{10} \rm yr$). One can calculate the expected DTD of GW sources, $\Gamma$, from this channel numerically, specifically we compute equations (\ref{eq:Gamma_MW}) and (\ref{eq:Gamma_ellipti}) for different integration times, $T_{\rm int}$.  If we ignore the ionization process, we expect a constant DTD, because the merger probability from equations (\ref{eq:empty_prop_approx}) and (\ref{eq:full_prob_apx_express}) are linear in $t$. However, in reality wide systems are continuously ionized in the field due to the flyby interactions, and there are less available binaries to merger via GW. The time dependence is given by Equations (\ref{eq:ionization empty}) and (\ref{eq:ionization full}).  We solve these equations numerically for different times after formation. We note that the total rate also depends on the SMA and local stellar density, which make the analytical dependence of $t$ very complicated,  therefore we only find the DTD numerically.  In Figure \ref{DTD} we present the DTD for MW-like galaxy, we present the best power law fit in order to compare to the classical $1/t$ we expect from SNe \citep{maoz2014} and sGRB \citep{Berger2014}.  We find for a MW-like galaxy the DTD is proportional to $\sim t^{-0.41}$. The exact value of the power law depends on the host galaxy characteristics (velocity dispersion and the distribution of the stellar density), as well as the distribution of the SMA of the wide binaries. 

\subsubsection{Mass ratio}

\begin{figure} 
\includegraphics[width=1\columnwidth]{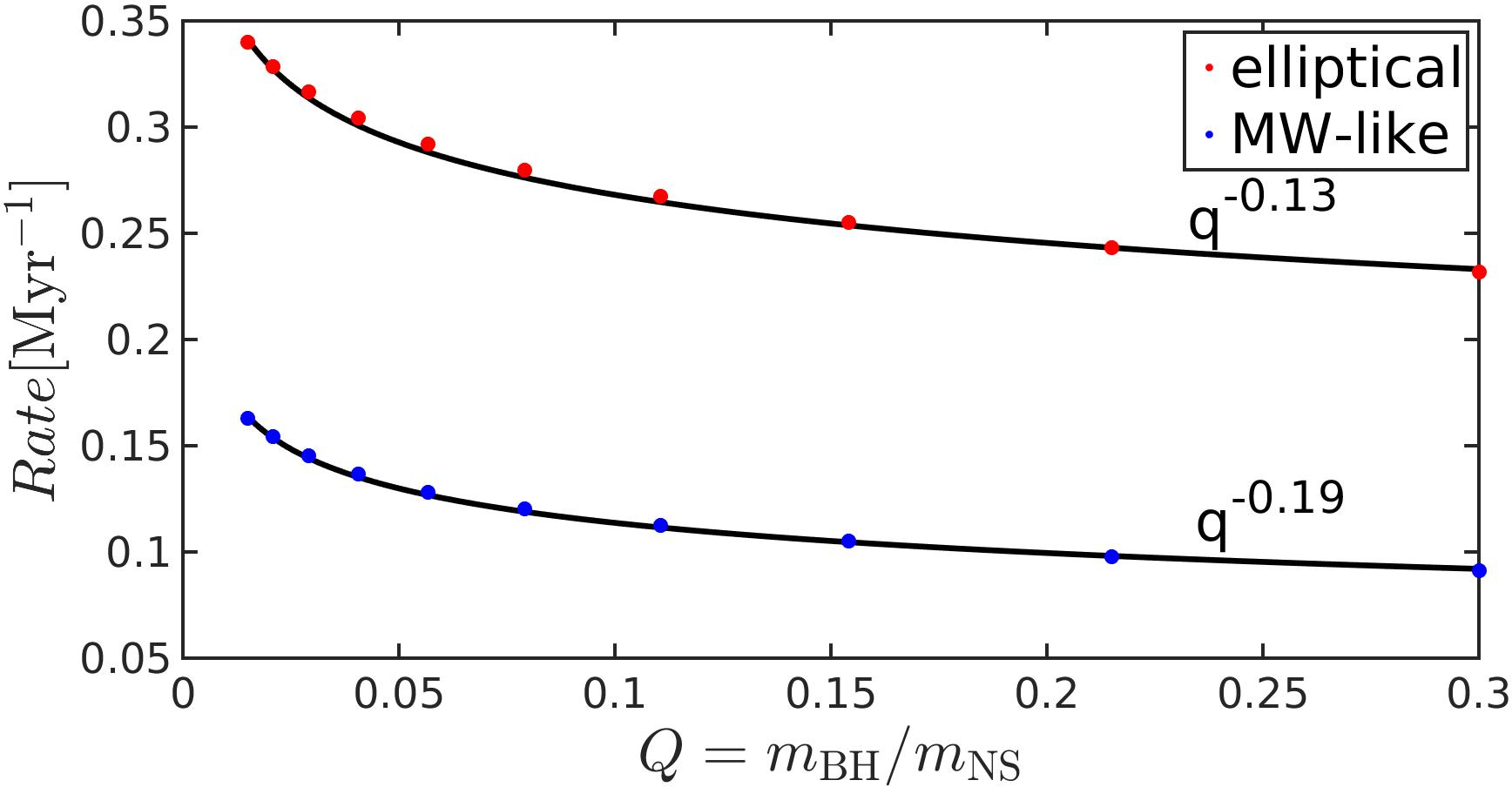} 
\caption{\label{Rate_Q} Average merger rate for a MW galaxy-like (red dots) and elliptical galaxy (blue dots). The velocity dispersion is $\sigma=40 \rm kms^{-1}$ for the MW-like galaxy and $\sigma=160 \rm kms^{-1}$ for the elliptical. Rate is calculated after $10^{10}\rm yrs$ from star formation, for binaries with $m_{\rm {NS}} =1.5M_\odot$ and varying $m_{\rm{BH}}=10M_\odot$. } 
\end{figure}

Here we present the model dependency on the mass ratio, $Q\equiv m_{\rm NS} / m_{\rm BH}$. Unlike \citet{Michaely2019} that considered BBH, here one of the companions is a NS and therefore has a constant mass. It is important to note that the merger rate dependency on $Q$ might be a source of confusion. In our case, NS-BH binary, a change is $Q$ infers also a change in the total binary mass $M$, due to the constant NS mass. Moreover, in what follows we do not assume the BH mass distribution but rather just present the merger rate assuming all BHs have the same mass. In Figure \ref{Rate_Q} we present the merger rate as a function of mass ration $Q$. We find that for the BHNS case a higher merger rate for lower $Q$, as can seen in Figure \ref{Rate_Q}.

\subsubsection{Spin distribution}
Because the origin of these mergers are from extremely wide binaries we do not expect these compact objects to have any correlation in their respective spins. Therefore, one should not expect any spin alignment between the BH and the NS in the moment of merger. Similar to other dynamically induced mergers the $\xi_{{\rm eff}}$
is expected to be close to zero. 


\subsection{sGRB}

Any BHNS merger is a potential for sGRB \citep{Berger2014}. A merger will not result in a disruption of the NS if the NS enters the BH event horizon in its entirety and therefore EM radiation is not created and omitted. The physical quantity that governs the outcome of the merger, whether or not it will produce an EM transient is the NS equation of state (EOS) which is still unconstrained. \cite{Fragione2021} showed that for the fraction of BH-NS
mergers that produce a sGRB, under some assumptions of their EOS is $ \sim1-50\%$. Accounting for this fraction and our merger rates we get that the sGRB from $R_{\rm {tot}}$:
\begin{equation}
    \Gamma_{\rm {sGRB}} \approx 0.5-20 \rm {Gpc^{3}yr^{-1}}.
\end{equation}
Additionally, unlike GW mergers detection, which we have very poor localization, and currently cannot assign a host galaxy to the detection, we are able in principle to identify the host galaxy, and check this model assumptions. For example, as mentioned above, sGRB are observed in both star forming galaxies and in ellipticals. This can be explained when considering the DTD of this model, see Figure \ref{DTD}. The DTD extends from star formation till $10^{10} \rm{yr}$, thus we expect these mergers and sGRBs to in both star forming and elliptical galaxies.

\subsection{Summary}
In this paper we present a novel dynamical channel for the formation of field wide BH-NS binaries mergers via GWs by random flyby interactions with passing stars. We find that the volumetric merger rate is up to $R_{\rm tot}\approx 40\rm Gpc^{3}yr^{-1}$.
Additionally, we report the following signatures:
(1) the merger rate increase with the host galaxy velocity dispersion.

(2) The DTD is function of the galaxy characteristics and shallower than uniform but steeper that $1/t$. For a MW-like spiral galaxy is $\propto t^{-0.41}$.

(3) The merger rate decrease with increasing mass ration of the binary.

(4) The effective spin distribution is expected to be zero due to the fact that the merger is originating from initially wide binary system, which we do not expect to have any correlations in a the spins.

(5) EM radiation, in the form of sGRB is expected to occur from $1-50\%$ of the merger we calculate, which translates in for this scenario up to $0.4-20 \rm{Gpc^3yr^{-1}}$.

\section*{Acknowledgments}
 EM and SN thank the Bhaumik Institute for Theoretical Physics and Howard and Astrid Preston for their generous support. We also acknowledges the partial support from NASA ATP 80NSSC20K0505. 
\appendix
Here we present the explicit function form of the rate equation. We distinguish between two different regimes, the impulsive regime and the kick regime.

\textbf{Impulsive regime.} In this case $b_{\rm impulsive}<b_{\rm kick}$. We remind the reader that $b_{\rm impulsive} = \alpha P v_{\rm enc}$, hence the explicit form of the critical SMA is
\begin{equation}
    a_{\rm crit}=\left(\frac{\left(GM\right)^{3/2}}{8\pi^4n_*v_{\rm enc}^3\alpha^2}\right)^{2/9}
\end{equation} 
and the corresponding $f$ is
\begin{equation}
    f(a)=n_*\alpha^2P(a)^2v_{\rm enc}^3=n_*\alpha^2\left(\frac{4\pi^2a^3}{GM}\right)v_{\rm enc}^3.
\end{equation}
The loss cone $F_q$ is given by (\ref{eq:F_q}), hence the loss rate function for the impulsive regime, without accounting for binary ionization, is given by the following two equations, the first for the empty loss cone regime:
\begin{equation}
    L_{a<a_{\rm crit}}=tfF_q= tn_*\alpha^2\left(\frac{4\pi^2a^3}{GM}\right)v_{\rm enc}^3\left(\frac{\beta}{f(a)a^4}\right)^{2/7}\propto a
\end{equation}
and for the full loss cone regime
\begin{equation}
    L_{a>a_{\rm crit}}=t\frac{F_q}{P(a)}= t \left(\frac{\beta}{f(a)a^4}\right)^{2/7}\left(\frac{GM}{4\pi^2a^3}\right)\propto a^{-5}.
\end{equation}
However, in order to calculate the rate including the ionization that the binaries experience in the field one should take into account that half lifetime of the binaries from equation (\ref{eq:t_half_life}). This introduces a nontrivial dependency on the SMA, $a$ and we get 
\begin{equation}
    L_{a<a_{\rm crit}}=\tau fF_q\left(1-e^{-t/\tau}\right)= \tau n_*\alpha^2\left(\frac{4\pi^2a^3}{GM}\right)v_{\rm enc}^3\left(\frac{\beta}{f(a)a^4}\right)^{2/7}\left(1-e^{-t/\tau}\right)
\end{equation}
and for the full loss cone
\begin{equation}
    L_{a>a_{\rm crit}}=\tau\frac{F_q}{P(a)}\left(1-e^{-t/\tau}\right)= \tau \left(\frac{\beta}{f(a)a^4}\right)^{2/7}\left(\frac{GM}{4\pi^2a^3}\right)\left(1-e^{-t/\tau}\right).
\end{equation}
\textbf{Kick regime}. In this case $b_{\rm impulsive}>b_{\rm kick}$. In this case $b_{\rm kick}$ is given by equation (\ref{eq:b_kick}), and the corresponding $a_{\rm crit}$ and $f$ are given by
\begin{equation}
    a_{\rm crit}=\left(\frac{GM}{n_*^2\pi^2}\right)^{6/47}\left(\frac{4M}{27Gm_p^2}\left(\frac{\beta}{n_*\pi}\right)^{2/7}\right)^{7/47}
\end{equation} 
and
\begin{equation}
    f=n_*\sigma(b_{\rm kick})v_{\rm enc}=n_*\pi\left(\frac{27G}{4} a^{29/7} \frac{m_p^2}{M} \left( \frac{n_*\pi}{\beta'} \right)^{2/7}\right)^{7/12}.
\end{equation}
Next we write the rate equations for both the empty and full cone regimes, without binary ionization to get
\begin{equation}
    L_{a<a_{\rm crit}}=tfF_q= tn_*\pi\left(\frac{27G}{4} a^{29/7} \frac{m_p^2}{M} \left( \frac{n_*\pi}{\beta'} \right)^{2/7}\right)^{7/12}\left(\frac{\beta}{f(a)a^4}\right)^{2/7}\propto a^{7/12}
\end{equation}
and
\begin{equation}
    L_{a>a_{\rm crit}}=t\frac{F_q}{P(a)}= t \left(\frac{\beta}{f(a)a^4}\right)^{2/7}\left(\frac{GM}{4\pi^2a^3}\right)\propto a^{-29/6}.
\end{equation}
Next we write the same rate equation with the correction for the binary ionization:
\begin{equation}
    L_{a<a_{\rm crit}}= \tau n_*\pi\left(\frac{27G}{4} a^{29/7} \frac{m_p^2}{M} \left( \frac{n_*\pi}{\beta'} \right)^{2/7}\right)^{7/12}\left(\frac{\beta}{f(a)a^4}\right)^{2/7}\left(1-e^{-t/\tau}\right)
\end{equation}
and
\begin{equation}
    L_{a>a_{\rm crit}}=t\frac{F_q}{P(a)}= \tau \left(\frac{\beta}{f(a)a^4}\right)^{2/7}\left(\frac{GM}{4\pi^2a^3}\right)\left(1-e^{-t/\tau}\right).
\end{equation}

\bibliographystyle{aasjournal}

\begin{thebibliography}{}
\expandafter\ifx\csname natexlab\endcsname\relax\def\natexlab#1{#1}\fi
\providecommand{\url}[1]{\href{#1}{#1}}
\providecommand{\dodoi}[1]{doi:~\href{http://doi.org/#1}{\nolinkurl{#1}}}
\providecommand{\doeprint}[1]{\href{http://ascl.net/#1}{\nolinkurl{http://ascl.net/#1}}}
\providecommand{\doarXiv}[1]{\href{https://arxiv.org/abs/#1}{\nolinkurl{https://arxiv.org/abs/#1}}}

\bibitem[{{Antonini} {et~al.}(2016){Antonini}, {Chatterjee}, {Rodriguez},
  {Morscher}, {Pattabiraman}, {Kalogera}, \& {Rasio}}]{Antonini2016}
{Antonini}, F., {Chatterjee}, S., {Rodriguez}, C.~L., {et~al.} 2016, \apj, 816,
  65, \dodoi{10.3847/0004-637X/816/2/65}

\bibitem[{{Antonini} {et~al.}(2014){Antonini}, {Murray}, \&
  {Mikkola}}]{Antonini2014}
{Antonini}, F., {Murray}, N., \& {Mikkola}, S. 2014, \apj, 781, 45,
  \dodoi{10.1088/0004-637X/781/1/45}

\bibitem[{{Antonini} \& {Perets}(2012)}]{antonini2012}
{Antonini}, F., \& {Perets}, H.~B. 2012, \apj, 757, 27,
  \dodoi{10.1088/0004-637X/757/1/27}

\bibitem[{{Antonini} {et~al.}(2017){Antonini}, {Toonen}, \&
  {Hamers}}]{Antonini2017}
{Antonini}, F., {Toonen}, S., \& {Hamers}, A.~S. 2017, \apj, 841, 77,
  \dodoi{10.3847/1538-4357/aa6f5e}

\bibitem[{{Baes} {et~al.}(2007){Baes}, {Sil'chenko}, {Moiseev}, \&
  {Manakova}}]{Baes2007}
{Baes}, M., {Sil'chenko}, O.~K., {Moiseev}, A.~V., \& {Manakova}, E.~A. 2007,
  \aap, 467, 991, \dodoi{10.1051/0004-6361:20066758}

\bibitem[{{Bahcall} {et~al.}(1985){Bahcall}, {Hut}, \&
  {Tremaine}}]{Bahcall1985}
{Bahcall}, J.~N., {Hut}, P., \& {Tremaine}, S. 1985, \apj, 290, 15,
  \dodoi{10.1086/162953}

\bibitem[{{Banerjee}(2018)}]{Banerjee2018}
{Banerjee}, S. 2018, \mnras, 473, 909, \dodoi{10.1093/mnras/stx2347}

\bibitem[{{Bartos} {et~al.}(2017){Bartos}, {Haiman}, {Marka}, {Metzger},
  {Stone}, \& {Marka}}]{Bartos2017}
{Bartos}, I., {Haiman}, Z., {Marka}, Z., {et~al.} 2017, Nature Communications,
  8, 831, \dodoi{10.1038/s41467-017-00851-7}

\bibitem[{{Belczynski} {et~al.}(2008){Belczynski}, {Kalogera}, {Rasio}, {Taam},
  {Zezas}, {Bulik}, {Maccarone}, \& {Ivanova}}]{belczynski2008}
{Belczynski}, K., {Kalogera}, V., {Rasio}, F.~A., {et~al.} 2008, \apjs, 174,
  223, \dodoi{10.1086/521026}

\bibitem[{{Belczynski} {et~al.}(2016){Belczynski}, {Repetto}, {Holz},
  {O'Shaughnessy}, {Bulik}, {Berti}, {Fryer}, \& {Dominik}}]{Belczynski2016}
{Belczynski}, K., {Repetto}, S., {Holz}, D.~E., {et~al.} 2016, \apj, 819, 108,
  \dodoi{10.3847/0004-637X/819/2/108}

\bibitem[{{Belczynski} {et~al.}(2007){Belczynski}, {Taam}, {Kalogera}, {Rasio},
  \& {Bulik}}]{belczynski2007}
{Belczynski}, K., {Taam}, R.~E., {Kalogera}, V., {Rasio}, F.~A., \& {Bulik}, T.
  2007, \apj, 662, 504, \dodoi{10.1086/513562}

\bibitem[{{Berger}(2014)}]{Berger2014}
{Berger}, E. 2014, \araa, 52, 43, \dodoi{10.1146/annurev-astro-081913-035926}

\bibitem[{{Britt} {et~al.}(2021){Britt}, {Johanson}, {Wood}, {Miller}, \&
  {Michaely}}]{Britt2021}
{Britt}, D., {Johanson}, B., {Wood}, L., {Miller}, M.~C., \& {Michaely}, E.
  2021, \mnras, 505, 3844, \dodoi{10.1093/mnras/stab1570}

\bibitem[{{de Mink} \& {Belczynski}(2015)}]{deMink2015}
{de Mink}, S.~E., \& {Belczynski}, K. 2015, \apj, 814, 58,
  \dodoi{10.1088/0004-637X/814/1/58}

\bibitem[{{de Mink} \& {Mandel}(2016)}]{deMink2016}
{de Mink}, S.~E., \& {Mandel}, I. 2016, \mnras, 460, 3545,
  \dodoi{10.1093/mnras/stw1219}

\bibitem[{{Dominik} {et~al.}(2015){Dominik}, {Berti}, {O'Shaughnessy},
  {Mandel}, {Belczynski}, {Fryer}, {Holz}, {Bulik}, \&
  {Pannarale}}]{Dominik2015}
{Dominik}, M., {Berti}, E., {O'Shaughnessy}, R., {et~al.} 2015, \apj, 806, 263,
  \dodoi{10.1088/0004-637X/806/2/263}

\bibitem[{{Duch{\^e}ne} \& {Kraus}(2013)}]{Duchene2013}
{Duch{\^e}ne}, G., \& {Kraus}, A. 2013, \araa, 51, 269,
  \dodoi{10.1146/annurev-astro-081710-102602}

\bibitem[{{Eldridge} {et~al.}(2017){Eldridge}, {Stanway}, {Xiao}, {McClelland},
  {Taylor}, {Ng}, {Greis}, \& {Bray}}]{Eldridge2017}
{Eldridge}, J.~J., {Stanway}, E.~R., {Xiao}, L., {et~al.} 2017, \pasa, 34,
  e058, \dodoi{10.1017/pasa.2017.51}

\bibitem[{{Ertl} {et~al.}(2015){Ertl}, {Janka}, {Woosley}, {Sukhbold}, \&
  {Ugliano}}]{ertl2015}
{Ertl}, T., {Janka}, H.-T., {Woosley}, S.~E., {Sukhbold}, T., \& {Ugliano}, M.
  2015, ArXiv e-prints.
\newblock \doarXiv{1503.07522}

\bibitem[{{Fragione}(2021)}]{Fragione2021}
{Fragione}, G. 2021, \apjl, 923, L2, \dodoi{10.3847/2041-8213/ac3bcd}

\bibitem[{{Fragione} {et~al.}(2019){Fragione}, {Grishin}, {Leigh}, {Perets}, \&
  {Perna}}]{Fragione2019}
{Fragione}, G., {Grishin}, E., {Leigh}, N. W.~C., {Perets}, H.~B., \& {Perna},
  R. 2019, \mnras, 488, 47, \dodoi{10.1093/mnras/stz1651}

\bibitem[{{Fragione} \& {Kocsis}(2018)}]{Fragione2018}
{Fragione}, G., \& {Kocsis}, B. 2018, \prl, 121, 161103,
  \dodoi{10.1103/PhysRevLett.121.161103}

\bibitem[{{Fryer} {et~al.}(2012){Fryer}, {Belczynski}, {Wiktorowicz},
  {Dominik}, {Kalogera}, \& {Holz}}]{Fryer2012}
{Fryer}, C.~L., {Belczynski}, K., {Wiktorowicz}, G., {et~al.} 2012, \apj, 749,
  91, \dodoi{10.1088/0004-637X/749/1/91}

\bibitem[{{Gallegos-Garcia} {et~al.}(2021){Gallegos-Garcia}, {Berry},
  {Marchant}, \& {Kalogera}}]{GallegosGarcia2021}
{Gallegos-Garcia}, M., {Berry}, C. P.~L., {Marchant}, P., \& {Kalogera}, V.
  2021, \apj, 922, 110, \dodoi{10.3847/1538-4357/ac2610}

\bibitem[{{Giacobbo} {et~al.}(2018){Giacobbo}, {Mapelli}, \&
  {Spera}}]{Giacobbo2018}
{Giacobbo}, N., {Mapelli}, M., \& {Spera}, M. 2018, \mnras, 474, 2959,
  \dodoi{10.1093/mnras/stx2933}

\bibitem[{{Hamers} {et~al.}(2018){Hamers}, {Bar-Or}, {Petrovich}, \&
  {Antonini}}]{Hamers2018}
{Hamers}, A.~S., {Bar-Or}, B., {Petrovich}, C., \& {Antonini}, F. 2018, \apj,
  865, 2, \dodoi{10.3847/1538-4357/aadae2}

\bibitem[{{Hamilton} \& {Rafikov}(2019)}]{Hamilton2019}
{Hamilton}, C., \& {Rafikov}, R.~R. 2019, \mnras, 488, 5512,
  \dodoi{10.1093/mnras/stz2026}

\bibitem[{{Heggie}(1975)}]{heggie1975}
{Heggie}, D.~C. 1975, \mnras, 173, 729, \dodoi{10.1093/mnras/173.3.729}

\bibitem[{{Hernquist}(1990)}]{hernquist1990}
{Hernquist}, L. 1990, \apj, 356, 359, \dodoi{10.1086/168845}

\bibitem[{{Hills}(1981)}]{Hills1981}
{Hills}, J.~G. 1981, \aj, 86, 1730, \dodoi{10.1086/113058}

\bibitem[{{Hoang} {et~al.}(2018){Hoang}, {Naoz}, {Kocsis}, {Rasio}, \&
  {Dosopoulou}}]{Hoang2018}
{Hoang}, B.-M., {Naoz}, S., {Kocsis}, B., {Rasio}, F.~A., \& {Dosopoulou}, F.
  2018, \apj, 856, 140, \dodoi{10.3847/1538-4357/aaafce}

\bibitem[{{Hoang} {et~al.}(2020){Hoang}, {Naoz}, \& {Kremer}}]{Hoang2020}
{Hoang}, B.-M., {Naoz}, S., \& {Kremer}, K. 2020, \apj, 903, 8,
  \dodoi{10.3847/1538-4357/abb66a}

\bibitem[{{Juri{\'c}} {et~al.}(2008){Juri{\'c}}, {Ivezi{\'c}}, {Brooks},
  {Lupton}, {Schlegel}, {Finkbeiner}, {Padmanabhan}, {Bond}, {Sesar},
  {Rockosi}, {Knapp}, {Gunn}, {Sumi}, {Schneider}, {Barentine}, {Brewington},
  {Brinkmann}, {Fukugita}, {Harvanek}, {Kleinman}, {Krzesinski}, {Long},
  {Neilsen}, {Nitta}, {Snedden}, \& {York}}]{Juric2008}
{Juri{\'c}}, M., {Ivezi{\'c}}, {\v Z}., {Brooks}, A., {et~al.} 2008, \apj, 673,
  864, \dodoi{10.1086/523619}

\bibitem[{{Kaib} \& {Raymond}(2014)}]{kaib2014}
{Kaib}, N.~A., \& {Raymond}, S.~N. 2014, \apj, 782, 60,
  \dodoi{10.1088/0004-637X/782/2/60}

\bibitem[{{Kimpson} {et~al.}(2016){Kimpson}, {Spera}, {Mapelli}, \&
  {Ziosi}}]{Kimpson2016}
{Kimpson}, T.~O., {Spera}, M., {Mapelli}, M., \& {Ziosi}, B.~M. 2016, \mnras,
  463, 2443, \dodoi{10.1093/mnras/stw2085}

\bibitem[{{Kopparapu} {et~al.}(2008){Kopparapu}, {Hanna}, {Kalogera},
  {O'Shaughnessy}, {Gonz{\'a}lez}, {Brady}, \& {Fairhurst}}]{Kopparapu2008}
{Kopparapu}, R.~K., {Hanna}, C., {Kalogera}, V., {et~al.} 2008, \apj, 675,
  1459, \dodoi{10.1086/527348}

\bibitem[{{Kozai}(1962)}]{Kozai1962}
{Kozai}, Y. 1962, \aj, 67, 591

\bibitem[{{Kremer} {et~al.}(2020){Kremer}, {Spera}, {Becker}, {Chatterjee}, {Di
  Carlo}, {Fragione}, {Rodriguez}, {Ye}, \& {Rasio}}]{Kremer2020}
{Kremer}, K., {Spera}, M., {Becker}, D., {et~al.} 2020, \apj, 903, 45,
  \dodoi{10.3847/1538-4357/abb945}

\bibitem[{{Kroupa}(2001)}]{Kroupa2001}
{Kroupa}, P. 2001, \mnras, 322, 231, \dodoi{10.1046/j.1365-8711.2001.04022.x}

\bibitem[{{Leibler} \& {Berger}(2010)}]{Leibler2010}
{Leibler}, C.~N., \& {Berger}, E. 2010, \apj, 725, 1202,
  \dodoi{10.1088/0004-637X/725/1/1202}

\bibitem[{{Leigh} {et~al.}(2018){Leigh}, {Geller}, {McKernan}, {Ford}, {Mac
  Low}, {Bellovary}, {Haiman}, {Lyra}, {Samsing}, {O'Dowd}, {Kocsis}, \&
  {Endlich}}]{Leigh2018}
{Leigh}, N.~W.~C., {Geller}, A.~M., {McKernan}, B., {et~al.} 2018, \mnras, 474,
  5672, \dodoi{10.1093/mnras/stx3134}

\bibitem[{{Lidov}(1962)}]{Lidov1962}
{Lidov}, M.~L. 1962, \planss, 9, 719, \dodoi{10.1016/0032-0633(62)90129-0}

\bibitem[{{Lightman} \& {Shapiro}(1977)}]{Lightman1977}
{Lightman}, A.~P., \& {Shapiro}, S.~L. 1977, \apj, 211, 244,
  \dodoi{10.1086/154925}

\bibitem[{{Liu} \& {Lai}(2018)}]{Liu2018}
{Liu}, B., \& {Lai}, D. 2018, \apj, 863, 68, \dodoi{10.3847/1538-4357/aad09f}

\bibitem[{{Mandel}(2016)}]{Mandel2016a}
{Mandel}, I. 2016, \mnras, 456, 578, \dodoi{10.1093/mnras/stv2733}

\bibitem[{{Mandel} \& {de Mink}(2016)}]{Mandel2016}
{Mandel}, I., \& {de Mink}, S.~E. 2016, \mnras, 458, 2634,
  \dodoi{10.1093/mnras/stw379}

\bibitem[{{Mandel} \& {Smith}(2021)}]{Mandel2021}
{Mandel}, I., \& {Smith}, R. J.~E. 2021, \apjl, 922, L14,
  \dodoi{10.3847/2041-8213/ac35dd}

\bibitem[{{Maoz} {et~al.}(2014){Maoz}, {Mannucci}, \& {Nelemans}}]{maoz2014}
{Maoz}, D., {Mannucci}, F., \& {Nelemans}, G. 2014, \araa, 52, 107,
  \dodoi{10.1146/annurev-astro-082812-141031}

\bibitem[{{Marchant} {et~al.}(2016){Marchant}, {Langer}, {Podsiadlowski},
  {Tauris}, \& {Moriya}}]{Marchant2016}
{Marchant}, P., {Langer}, N., {Podsiadlowski}, P., {Tauris}, T.~M., \&
  {Moriya}, T.~J. 2016, \aap, 588, A50, \dodoi{10.1051/0004-6361/201628133}

\bibitem[{{Martinez} {et~al.}(2020){Martinez}, {Fragione}, {Kremer},
  {Chatterjee}, {Rodriguez}, {Samsing}, {Ye}, {Weatherford}, {Zevin}, {Naoz},
  \& {Rasio}}]{Martinez2020}
{Martinez}, M. A.~S., {Fragione}, G., {Kremer}, K., {et~al.} 2020, \apj, 903,
  67, \dodoi{10.3847/1538-4357/abba25}

\bibitem[{{Merritt}(2013)}]{Merritt2013}
{Merritt}, D. 2013, Classical and Quantum Gravity, 30, 244005,
  \dodoi{10.1088/0264-9381/30/24/244005}

\bibitem[{{Michaely}(2020)}]{Michaely2020b}
{Michaely}, E. 2020, \mnras, \dodoi{10.1093/mnras/staa3623}

\bibitem[{{Michaely} \& {Perets}(2016)}]{Michaely2016}
{Michaely}, E., \& {Perets}, H.~B. 2016, \mnras, 458, 4188,
  \dodoi{10.1093/mnras/stw368}

\bibitem[{{Michaely} \& {Perets}(2019)}]{Michaely2019}
---. 2019, \apjl, 887, L36, \dodoi{10.3847/2041-8213/ab5b9b}

\bibitem[{Michaely \& Perets(2020)}]{Michaely2020}
Michaely, E., \& Perets, H.~B. 2020, Monthly Notices of the Royal Astronomical
  Society, \dodoi{10.1093/mnras/staa2720}

\bibitem[{{Michaely} \& {Shara}(2021)}]{Michaely2021}
{Michaely}, E., \& {Shara}, M.~M. 2021, \mnras, 502, 4540,
  \dodoi{10.1093/mnras/stab339}

\bibitem[{{Miller} \& {Hamilton}(2002)}]{Miller2002a}
{Miller}, M.~C., \& {Hamilton}, D.~P. 2002, \apj, 576, 894,
  \dodoi{10.1086/341788}

\bibitem[{{Moe} \& {Di Stefano}(2016)}]{Moe2016}
{Moe}, M., \& {Di Stefano}, R. 2016, ArXiv e-prints.
\newblock \doarXiv{1606.05347}

\bibitem[{{Naoz}(2016)}]{Naoz2016}
{Naoz}, S. 2016, \araa, 54, 441, \dodoi{10.1146/annurev-astro-081915-023315}

\bibitem[{{Nomoto}(1982)}]{nomoto1982a}
{Nomoto}, K. 1982, \apj, 253, 798, \dodoi{10.1086/159682}

\bibitem[{{Nomoto}(1984)}]{nomoto1984}
---. 1984, \apj, 277, 791, \dodoi{10.1086/161749}

\bibitem[{{Olejak} {et~al.}(2020){Olejak}, {Fishbach}, {Belczynski}, {Holz},
  {Lasota}, {Miller}, \& {Bulik}}]{Olejak2020}
{Olejak}, A., {Fishbach}, M., {Belczynski}, K., {et~al.} 2020, \apjl, 901, L39,
  \dodoi{10.3847/2041-8213/abb5b5}

\bibitem[{{Peters}(1964)}]{Pet64}
{Peters}, P.~C. 1964, Physical Review, 136, 1224,
  \dodoi{10.1103/PhysRev.136.B1224}

\bibitem[{{Petrovich} \& {Antonini}(2017)}]{Petrovich2017}
{Petrovich}, C., \& {Antonini}, F. 2017, \apj, 846, 146,
  \dodoi{10.3847/1538-4357/aa8628}

\bibitem[{{Rastello} {et~al.}(2021){Rastello}, {Mapelli}, {Di Carlo}, {Iorio},
  {Ballone}, {Giacobbo}, {Santoliquido}, \& {Torniamenti}}]{Rastello2021}
{Rastello}, S., {Mapelli}, M., {Di Carlo}, U.~N., {et~al.} 2021, \mnras, 507,
  3612, \dodoi{10.1093/mnras/stab2355}

\bibitem[{{Repetto} {et~al.}(2012){Repetto}, {Davies}, \&
  {Sigurdsson}}]{Repetto2012}
{Repetto}, S., {Davies}, M.~B., \& {Sigurdsson}, S. 2012, \mnras, 425, 2799,
  \dodoi{10.1111/j.1365-2966.2012.21549.x}

\bibitem[{{Repetto} {et~al.}(2017){Repetto}, {Igoshev}, \&
  {Nelemans}}]{Repetto2017}
{Repetto}, S., {Igoshev}, A.~P., \& {Nelemans}, G. 2017, \mnras, 467, 298,
  \dodoi{10.1093/mnras/stx027}

\bibitem[{{Rodriguez} {et~al.}(2018){Rodriguez}, {Amaro-Seoane}, {Chatterjee},
  {Kremer}, {Rasio}, {Samsing}, {Ye}, \& {Zevin}}]{Rodriguez2018}
{Rodriguez}, C.~L., {Amaro-Seoane}, P., {Chatterjee}, S., {et~al.} 2018, \prd,
  98, 123005, \dodoi{10.1103/PhysRevD.98.123005}

\bibitem[{{Rodriguez} {et~al.}(2016){Rodriguez}, {Chatterjee}, \&
  {Rasio}}]{Rodriguez2016}
{Rodriguez}, C.~L., {Chatterjee}, S., \& {Rasio}, F.~A. 2016, \prd, 93, 084029,
  \dodoi{10.1103/PhysRevD.93.084029}

\bibitem[{{Rodriguez} {et~al.}(2021){Rodriguez}, {Kremer}, {Chatterjee},
  {Fragione}, {Loeb}, {Rasio}, {Weatherford}, \& {Ye}}]{Rodriguez2021}
{Rodriguez}, C.~L., {Kremer}, K., {Chatterjee}, S., {et~al.} 2021, Research
  Notes of the American Astronomical Society, 5, 19,
  \dodoi{10.3847/2515-5172/abdf54}

\bibitem[{{Salpeter}(1955)}]{salpeter1955}
{Salpeter}, E.~E. 1955, \apj, 121, 161, \dodoi{10.1086/145971}

\bibitem[{{Samsing} {et~al.}(2014){Samsing}, {MacLeod}, \&
  {Ramirez-Ruiz}}]{Samsing2014}
{Samsing}, J., {MacLeod}, M., \& {Ramirez-Ruiz}, E. 2014, \apj, 784, 71,
  \dodoi{10.1088/0004-637X/784/1/71}

\bibitem[{{Samsing} {et~al.}(2018){Samsing}, {MacLeod}, \&
  {Ramirez-Ruiz}}]{Samsing2018}
---. 2018, \apj, 853, 140, \dodoi{10.3847/1538-4357/aaa715}

\bibitem[{{Silsbee} \& {Tremaine}(2017)}]{Silsbee2017}
{Silsbee}, K., \& {Tremaine}, S. 2017, \apj, 836, 39,
  \dodoi{10.3847/1538-4357/aa5729}

\bibitem[{{Stephan} {et~al.}(2016){Stephan}, {Naoz}, {Ghez}, {Witzel},
  {Sitarski}, {Do}, \& {Kocsis}}]{Stephan2016}
{Stephan}, A.~P., {Naoz}, S., {Ghez}, A.~M., {et~al.} 2016, \mnras, 460, 3494,
  \dodoi{10.1093/mnras/stw1220}

\bibitem[{{Stephan} {et~al.}(2019){Stephan}, {Naoz}, {Ghez}, {Morris},
  {Ciurlo}, {Do}, {Breivik}, {Coughlin}, \& {Rodriguez}}]{Stephan2019}
---. 2019, \apj, 878, 58, \dodoi{10.3847/1538-4357/ab1e4d}

\bibitem[{{Stone} {et~al.}(2017){Stone}, {Metzger}, \& {Haiman}}]{Stone2017}
{Stone}, N.~C., {Metzger}, B.~D., \& {Haiman}, Z. 2017, \mnras, 464, 946,
  \dodoi{10.1093/mnras/stw2260}

\bibitem[{{The LIGO Scientific Collaboration} {et~al.}(2021{\natexlab{a}}){The
  LIGO Scientific Collaboration}, {the Virgo Collaboration}, {the KAGRA
  Collaboration}, {Abbott}, {Abbott}, {Acernese}, {Ackley}, {Adams},
  {Adhikari}, {Adhikari}, {Adya}, {Affeldt}, {Agarwal}, {Agathos}, {Agatsuma},
  {Aggarwal}, {Aguiar}, {Aiello}, {Ain}, {Ajith}, {Akcay}, {Akutsu},
  {Albanesi}, {Allocca}, {Altin}, {Amato}, {Anand}, {Anand}, {Ananyeva},
  {Anderson}, {Anderson}, {Ando}, {Andrade}, {Andres}, {Andri{\'c}},
  {Angelova}, {Ansoldi}, {Antelis}, {Antier}, {Appert}, {Arai}, {Arai}, {Arai},
  {Araki}, {Araya}, {Araya}, {Areeda}, {Ar{\`e}ne}, {Aritomi}, {Arnaud},
  {Arogeti}, {Aronson}, {Arun}, {Asada}, {Asali}, {Ashton}, {Aso}, {Assiduo},
  {Aston}, {Astone}, {Aubin}, {Austin}, {Babak}, {Badaracco}, {Bader},
  {Badger}, {Bae}, {Bae}, {Baer}, {Bagnasco}, {Bai}, {Baiotti}, {Baird},
  {Bajpai}, {Ball}, {Ballardin}, {Ballmer}, {Balsamo}, {Baltus}, {Banagiri},
  {Bankar}, {Barayoga}, {Barbieri}, {Barish}, {Barker}, {Barneo}, {Barone},
  {Barr}, {Barsotti}, {Barsuglia}, {Barta}, {Bartlett}, {Barton}, {Bartos},
  {Bassiri}, {Basti}, {Bawaj}, {Bayley}, {Baylor}, {Bazzan}, {B{\'e}csy},
  {Bedakihale}, {Bejger}, {Belahcene}, {Benedetto}, {Beniwal}, {Bennett},
  {Bentley}, {BenYaala}, {Bergamin}, {Berger}, {Bernuzzi}, {Berry},
  {Bersanetti}, {Bertolini}, {Betzwieser}, {Beveridge}, {Bhandare}, {Bhardwaj},
  {Bhattacharjee}, {Bhaumik}, {Bilenko}, {Billingsley}, {Bini}, {Birney},
  {Birnholtz}, {Biscans}, {Bischi}, {Biscoveanu}, {Bisht}, {Biswas}, {Bitossi},
  {Bizouard}, {Blackburn}, {Blair}, {Blair}, {Blair}, {Bobba}, {Bode}, {Boer},
  {Bogaert}, {Boldrini}, {Bonavena}, {Bondu}, {Bonilla}, {Bonnand}, {Booker},
  {Boom}, {Bork}, {Boschi}, {Bose}, {Bose}, {Bossilkov}, {Boudart},
  {Bouffanais}, {Bozzi}, {Bradaschia}, {Brady}, {Bramley}, {Branch},
  {Branchesi}, {Brandt}, {Brau}, {Breschi}, {Briant}, {Briggs}, {Brillet},
  {Brinkmann}, {Brockill}, {Brooks}, {Brooks}, {Brown}, {Brunett}, {Bruno},
  {Bruntz}, {Bryant}, {Bulik}, {Bulten}, {Buonanno}, {Buscicchio}, {Buskulic},
  {Buy}, {Byer}, {Cabourn Davies}, {Cadonati}, {Cagnoli}, {Cahillane},
  {Calder{\'o}n Bustillo}, {Callaghan}, {Callister}, {Calloni}, {Cameron},
  {Camp}, {Canepa}, {Canevarolo}, {Cannavacciuolo}, {Cannon}, {Cao}, {Cao},
  {Capocasa}, {Capote}, {Carapella}, {Carbognani}, {Carlin}, {Carney},
  {Carpinelli}, {Carrillo}, {Carullo}, {Carver}, {Casanueva Diaz}, {Casentini},
  {Castaldi}, {Caudill}, {Cavagli{\`a}}, {Cavalier}, {Cavalieri}, {Ceasar},
  {Cella}, {Cerd{\'a}-Dur{\'a}n}, {Cesarini}, {Chaibi}, {Chakravarti},
  {Chalathadka Subrahmanya}, {Champion}, {Chan}, {Chan}, {Chan}, {Chan},
  {Chan}, {Chandra}, {Chanial}, {Chao}, {Chapman-Bird}, {Charlton}, {Chase},
  {Chassande-Mottin}, {Chatterjee}, {Chatterjee}, {Chatterjee}, {Chaturvedi},
  {Chaty}, {Chatziioannou}, {Chen}, {Chen}, {Chen}, {Chen}, {Chen}, {Chen},
  {Chen}, {Chen}, {Cheng}, {Cheong}, {Cheung}, {Chia}, {Chiadini}, {Chiang},
  {Chiarini}, {Chierici}, {Chincarini}, {Chiofalo}, {Chiummo}, {Cho}, {Cho},
  {Choudhary}, {Choudhary}, {Christensen}, {Chu}, {Chu}, {Chu}, {Chua},
  {Chung}, {Ciani}, {Ciecielag}, {Cie{\'s}lar}, {Cifaldi}, {Ciobanu}, {Ciolfi},
  {Cipriano}, {Cirone}, {Clara}, {Clark}, {Clark}, {Clarke}, {Clearwater},
  {Clesse}, {Cleva}, {Coccia}, {Codazzo}, {Cohadon}, {Cohen}, {Cohen},
  {Colleoni}, {Collette}, {Colombo}, {Colpi}, {Compton}, {Constancio}, {Conti},
  {Cooper}, {Corban}, {Corbitt}, {Cordero-Carri{\'o}n}, {Corezzi}, {Corley},
  {Cornish}, {Corre}, {Corsi}, {Cortese}, {Costa}, {Cotesta}, {Coughlin},
  {Coulon}, {Countryman}, {Cousins}, {Couvares}, {Coward}, {Cowart}, {Coyne},
  {Coyne}, {Creighton}, {Creighton}, {Criswell}, {Croquette}, {Crowder},
  {Cudell}, {Cullen}, {Cumming}, {Cummings}, {Cunningham}, {Cuoco},
  {Cury{\l}o}, {Dabadie}, {Dal Canton}, {Dall'Osso}, {D{\'a}lya}, {Dana},
  {DaneshgaranBajastani}, {D'Angelo}, {Danila}, {Danilishin}, {D'Antonio},
  {Danzmann}, {Darsow-Fromm}, {Dasgupta}, {Datrier}, {Datta}, {Dattilo},
  {Dave}, {Davier}, {Davis}, {Davis}, {Daw}, {de Alarc{\'o}n}, {Dean}, {DeBra},
  {Deenadayalan}, {Degallaix}, {De Laurentis}, {Del{\'e}glise}, {Del Favero},
  {De Lillo}, {De Lillo}, {Del Pozzo}, {DeMarchi}, {De Matteis}, {D'Emilio},
  {Demos}, {Dent}, {Depasse}, {De Pietri}, {De Rosa}, {De Rossi}, {DeSalvo},
  {De Simone}, {Dhurandhar}, {D{\'\i}az}, {Diaz-Ortiz}, {Didio}, {Dietrich},
  {Di Fiore}, {Di Fronzo}, {Di Giorgio}, {Di Giovanni}, {Di Giovanni}, {Di
  Girolamo}, {Di Lieto}, {Ding}, {Di Pace}, {Di Palma}, {Di Renzo},
  {Divakarla}, {Dmitriev}, {Doctor}, {D'Onofrio}, {Donovan}, {Dooley},
  {Doravari}, {Dorrington}, {Drago}, {Driggers}, {Drori}, {Ducoin}, {Dupej},
  {Durante}, {D'Urso}, {Duverne}, {Dwyer}, {Eassa}, {Easter}, {Ebersold},
  {Eckhardt}, {Eddolls}, {Edelman}, {Edo}, {Edy}, {Effler}, {Eguchi},
  {Eichholz}, {Eikenberry}, {Eisenmann}, {Eisenstein}, {Ejlli}, {Engelby},
  {Enomoto}, {Errico}, {Essick}, {Estell{\'e}s}, {Estevez}, {Etienne}, {Etzel},
  {Evans}, {Evans}, {Ewing}, {Fafone}, {Fair}, {Fairhurst}, {Farah}, {Farinon},
  {Farr}, {Farr}, {Farrow}, {Fauchon-Jones}, {Favaro}, {Favata}, {Fays},
  {Fazio}, {Feicht}, {Fejer}, {Fenyvesi}, {Ferguson}, {Fernandez-Galiana},
  {Ferrante}, {Ferreira}, {Fidecaro}, {Figura}, {Fiori}, {Fishbach}, {Fisher},
  {Fittipaldi}, {Fiumara}, {Flaminio}, {Floden}, {Fong}, {Font}, {Fornal},
  {Forsyth}, {Franke}, {Frasca}, {Frasconi}, {Frederick}, {Freed}, {Frei},
  {Freise}, {Frey}, {Fritschel}, {Frolov}, {Fronz{\'e}}, {Fujii}, {Fujikawa},
  {Fukunaga}, {Fukushima}, {Fulda}, {Fyffe}, {Gabbard}, {Gabella}, {Gadre},
  {Gair}, {Gais}, {Galaudage}, {Gamba}, {Ganapathy}, {Ganguly}, {Gao},
  {Gaonkar}, {Garaventa}, {Garc{\'\i}a}, {Garc{\'\i}a-N{\'u}{\~n}ez},
  {Garc{\'\i}a-Quir{\'o}s}, {Garufi}, {Gateley}, {Gaudio}, {Gayathri}, {Ge},
  {Gemme}, {Gennai}, {George}, {George}, {Gerberding}, {Gergely}, {Gewecke},
  {Ghonge}, {Ghosh}, {Ghosh}, {Ghosh}, {Ghosh}, {Giacomazzo}, {Giacoppo},
  {Giaime}, {Giardina}, {Gibson}, {Gier}, {Giesler}, {Giri}, {Gissi},
  {Glanzer}, {Gleckl}, {Godwin}, {Goetz}, {Goetz}, {Gohlke}, {Golomb},
  {Goncharov}, {Gonz{\'a}lez}, {Gopakumar}, {Gosselin}, {Gouaty}, {Gould},
  {Grace}, {Grado}, {Granata}, {Granata}, {Grant}, {Gras}, {Grassia}, {Gray},
  {Gray}, {Greco}, {Green}, {Green}, {Gretarsson}, {Gretarsson}, {Griffith},
  {Griffiths}, {Griggs}, {Grignani}, {Grimaldi}, {Grimm}, {Grote}, {Grunewald},
  {Gruning}, {Guerra}, {Guidi}, {Guimaraes}, {Guix{\'e}}, {Gulati}, {Guo},
  {Guo}, {Gupta}, {Gupta}, {Gupta}, {Gustafson}, {Gustafson}, {Guzman}, {Ha},
  {Haegel}, {Hagiwara}, {Haino}, {Halim}, {Hall}, {Hamilton}, {Hammond}, {Han},
  {Haney}, {Hanks}, {Hanna}, {Hannam}, {Hannuksela}, {Hansen}, {Hansen},
  {Hanson}, {Harder}, {Hardwick}, {Haris}, {Harms}, {Harry}, {Harry},
  {Hartwig}, {Hasegawa}, {Haskell}, {Hasskew}, {Haster}, {Hattori}, {Haughian},
  {Hayakawa}, {Hayama}, {Hayes}, {Healy}, {Heidmann}, {Heidt}, {Heintze},
  {Heinze}, {Heinzel}, {Heitmann}, {Hellman}, {Hello}, {Helmling-Cornell},
  {Hemming}, {Hendry}, {Heng}, {Hennes}, {Hennig}, {Hennig}, {Hernandez},
  {Hernandez Vivanco}, {Heurs}, {Hild}, {Hill}, {Himemoto}, {Hines},
  {Hiranuma}, {Hirata}, {Hirose}, {Hochheim}, {Hofman}, {Hohmann}, {Holcomb},
  {Holland}, {Holley-Bockelmann}, {Hollows}, {Holmes}, {Holt}, {Holz}, {Hong},
  {Hopkins}, {Hough}, {Hourihane}, {Howell}, {Hoy}, {Hoyland}, {Hreibi},
  {Hsieh}, {Hsu}, {Huang}, {Huang}, {Huang}, {Huang}, {Huang}, {Huang},
  {H{\"u}bner}, {Huddart}, {Hughey}, {Hui}, {Hui}, {Husa}, {Huttner},
  {Huxford}, {Huynh-Dinh}, {Ide}, {Idzkowski}, {Iess}, {Ikenoue}, {Imam},
  {Inayoshi}, {Ingram}, {Inoue}, {Ioka}, {Isi}, {Isleif}, {Ito}, {Itoh},
  {Iyer}, {Izumi}, {JaberianHamedan}, {Jacqmin}, {Jadhav}, {Jadhav}, {James},
  {Jan}, {Jani}, {Janquart}, {Janssens}, {Janthalur}, {Jaranowski}, {Jariwala},
  {Jaume}, {Jenkins}, {Jenner}, {Jeon}, {Jeunon}, {Jia}, {Jin}, {Johns},
  {Johnson-McDaniel}, {Jones}, {Jones}, {Jones}, {Jones}, {Jones}, {Jonker},
  {Ju}, {Jung}, {Jung}, {Junker}, {Juste}, {Kaihotsu}, {Kajita}, {Kakizaki},
  {Kalaghatgi}, {Kalogera}, {Kamai}, {Kamiizumi}, {Kanda}, {Kandhasamy},
  {Kang}, {Kanner}, {Kao}, {Kapadia}, {Kapasi}, {Karat}, {Karathanasis},
  {Karki}, {Kashyap}, {Kasprzack}, {Kastaun}, {Katsanevas}, {Katsavounidis},
  {Katzman}, {Kaur}, {Kawabe}, {Kawaguchi}, {Kawai}, {Kawasaki},
  {K{\'e}f{\'e}lian}, {Keitel}, {Key}, {Khadka}, {Khalili}, {Khan}, {Khazanov},
  {Khetan}, {Khursheed}, {Kijbunchoo}, {Kim}, {Kim}, {Kim}, {Kim}, {Kim},
  {Kim}, {Kimball}, {Kimura}, {Kinley-Hanlon}, {Kirchhoff}, {Kissel}, {Kita},
  {Kitazawa}, {Kleybolte}, {Klimenko}, {Knee}, {Knowles}, {Knyazev}, {Koch},
  {Koekoek}, {Kojima}, {Kokeyama}, {Koley}, {Kolitsidou}, {Kolstein}, {Komori},
  {Kondrashov}, {Kong}, {Kontos}, {Koper}, {Korobko}, {Kotake}, {Kovalam},
  {Kozak}, {Kozakai}, {Kozu}, {Kringel}, {Krishnendu}, {Kr{\'o}lak}, {Kuehn},
  {Kuei}, {Kuijer}, {Kulkarni}, {Kumar}, {Kumar}, {Kumar}, {Kumar}, {Kume},
  {Kuns}, {Kuo}, {Kuo}, {Kuromiya}, {Kuroyanagi}, {Kusayanagi}, {Kuwahara},
  {Kwak}, {Lagabbe}, {Laghi}, {Lalande}, {Lam}, {Lamberts}, {Landry}, {Lane},
  {Lang}, {Lange}, {Lantz}, {La Rosa}, {Lartaux-Vollard}, {Lasky}, {Laxen},
  {Lazzarini}, {Lazzaro}, {Leaci}, {Leavey}, {Lecoeuche}, {Lee}, {Lee}, {Lee},
  {Lee}, {Lee}, {Lee}, {Lehmann}, {Lema{\^\i}tre}, {Leonardi}, {Leroy},
  {Letendre}, {Levesque}, {Levin}, {Leviton}, {Leyde}, {Li}, {Li}, {Li}, {Li},
  {Li}, {Li}, {Lin}, {Lin}, {Lin}, {Lin}, {Lin}, {Linde}, {Linker}, {Linley},
  {Littenberg}, {Liu}, {Liu}, {Liu}, {Liu}, {Llamas}, {Llorens-Monteagudo},
  {Lo}, {Lockwood}, {Loh}, {London}, {Longo}, {Lopez}, {Lopez Portilla},
  {Lorenzini}, {Loriette}, {Lormand}, {Losurdo}, {Lott}, {Lough}, {Lousto},
  {Lovelace}, {Lucaccioni}, {L{\"u}ck}, {Lumaca}, {Lundgren}, {Luo}, {Lynam},
  {Macas}, {MacInnis}, {Macleod}, {MacMillan}, {Macquet}, {Maga{\~n}a
  Hernandez}, {Magazz{\`u}}, {Magee}, {Maggiore}, {Magnozzi}, {Mahesh},
  {Majorana}, {Makarem}, {Maksimovic}, {Maliakal}, {Malik}, {Man}, {Mandic},
  {Mangano}, {Mango}, {Mansell}, {Manske}, {Mantovani}, {Mapelli},
  {Marchesoni}, {Marchio}, {Marion}, {Mark}, {M{\'a}rka}, {M{\'a}rka},
  {Markakis}, {Markosyan}, {Markowitz}, {Maros}, {Marquina}, {Marsat},
  {Martelli}, {Martin}, {Martin}, {Martinez}, {Martinez}, {Martinez},
  {Martinovic}, {Martynov}, {Marx}, {Masalehdan}, {Mason}, {Massera},
  {Masserot}, {Massinger}, {Masso-Reid}, {Mastrogiovanni}, {Matas},
  {Mateu-Lucena}, {Matichard}, {Matiushechkina}, {Mavalvala}, {McCann},
  {McCarthy}, {McClelland}, {McClincy}, {McCormick}, {McCuller}, {McGhee},
  {McGuire}, {McIsaac}, {McIver}, {McRae}, {McWilliams}, {Meacher}, {Mehmet},
  {Mehta}, {Meijer}, {Melatos}, {Melchor}, {Mendell}, {Menendez-Vazquez},
  {Menoni}, {Mercer}, {Mereni}, {Merfeld}, {Merilh}, {Merritt}, {Merzougui},
  {Meshkov}, {Messenger}, {Messick}, {Meyers}, {Meylahn}, {Mhaske}, {Miani},
  {Miao}, {Michaloliakos}, {Michel}, {Michimura}, {Middleton}, {Milano},
  {Miller}, {Miller}, {Miller}, {Millhouse}, {Mills}, {Milotti}, {Minazzoli},
  {Minenkov}, {Mio}, {Mir}, {Miravet-Ten{\'e}s}, {Mishra}, {Mishra}, {Mistry},
  {Mitra}, {Mitrofanov}, {Mitselmakher}, {Mittleman}, {Miyakawa}, {Miyamoto},
  {Miyazaki}, {Miyo}, {Miyoki}, {Mo}, {Modafferi}, {Moguel}, {Mogushi},
  {Mohapatra}, {Mohite}, {Molina}, {Molina-Ruiz}, {Mondin}, {Montani}, {Moore},
  {Moraru}, {Morawski}, {More}, {Moreno}, {Moreno}, {Mori}, {Morisaki},
  {Moriwaki}, {Morr{\'a}s}, {Mours}, {Mow-Lowry}, {Mozzon}, {Muciaccia},
  {Mukherjee}, {Mukherjee}, {Mukherjee}, {Mukherjee}, {Mukherjee}, {Mukund},
  {Mullavey}, {Munch}, {Mu{\~n}iz}, {Murray}, {Musenich}, {Muusse}, {Nadji},
  {Nagano}, {Nagano}, {Nagar}, {Nakamura}, {Nakano}, {Nakano}, {Nakashima},
  {Nakayama}, {Napolano}, {Nardecchia}, {Narikawa}, {Naticchioni}, {Nayak},
  {Nayak}, {Negishi}, {Neil}, {Neilson}, {Nelemans}, {Nelson}, {Nery},
  {Neubauer}, {Neunzert}, {Ng}, {Ng}, {Nguyen}, {Nguyen}, {Nguyen}, {Nguyen
  Quynh}, {Ni}, {Nichols}, {Nishizawa}, {Nissanke}, {Nitoglia}, {Nocera},
  {Norman}, {North}, {Nozaki}, {Nu{\~n}o Siles}, {Nuttall}, {Oberling},
  {O'Brien}, {Obuchi}, {O'Dell}, {Oelker}, {Ogaki}, {Oganesyan}, {Oh}, {Oh},
  {Oh}, {Ohashi}, {Ohishi}, {Ohkawa}, {Ohme}, {Ohta}, {Okada}, {Okutani},
  {Okutomi}, {Olivetto}, {Oohara}, {Ooi}, {Oram}, {O'Reilly}, {Ormiston},
  {Ormsby}, {Ortega}, {O'Shaughnessy}, {O'Shea}, {Oshino}, {Ossokine},
  {Osthelder}, {Otabe}, {Ottaway}, {Overmier}, {Pace}, {Pagano}, {Page},
  {Pagliaroli}, {Pai}, {Pai}, {Palamos}, {Palashov}, {Palomba}, {Pan}, {Pan},
  {Panda}, {Pang}, {Pang}, {Pankow}, {Pannarale}, {Pant}, {Panther},
  {Paoletti}, {Paoli}, {Paolone}, {Parisi}, {Park}, {Park}, {Parker},
  {Pascucci}, {Pasqualetti}, {Passaquieti}, {Passuello}, {Patel}, {Pathak},
  {Patricelli}, {Patron}, {Paul}, {Payne}, {Pedraza}, {Pegoraro}, {Pele},
  {Pe{\~n}a Arellano}, {Penn}, {Perego}, {Pereira}, {Pereira}, {Perez},
  {P{\'e}rigois}, {Perkins}, {Perreca}, {Perri{\`e}s}, {Petermann},
  {Petterson}, {Pfeiffer}, {Pham}, {Phukon}, {Piccinni}, {Pichot},
  {Piendibene}, {Piergiovanni}, {Pierini}, {Pierro}, {Pillant}, {Pillas},
  {Pilo}, {Pinard}, {Pinto}, {Pinto}, {Piotrzkowski}, {Piotrzkowski},
  {Pirello}, {Pitkin}, {Placidi}, {Planas}, {Plastino}, {Pluchar}, {Poggiani},
  {Polini}, {Pong}, {Ponrathnam}, {Popolizio}, {Porter}, {Poulton}, {Powell},
  {Pracchia}, {Pradier}, {Prajapati}, {Prasai}, {Prasanna}, {Pratten},
  {Principe}, {Prodi}, {Prokhorov}, {Prosposito}, {Prudenzi}, {Puecher},
  {Punturo}, {Puosi}, {Puppo}, {P{\"u}rrer}, {Qi}, {Quetschke},
  {Quitzow-James}, {Qutob}, {Raab}, {Raaijmakers}, {Radkins}, {Radulesco},
  {Raffai}, {Rail}, {Raja}, {Rajan}, {Ramirez}, {Ramirez}, {Ramos-Buades},
  {Rana}, {Rapagnani}, {Rapol}, {Ray}, {Raymond}, {Raza}, {Razzano}, {Read},
  {Rees}, {Regimbau}, {Rei}, {Reid}, {Reid}, {Reitze}, {Relton}, {Renzini},
  {Rettegno}, {Reza}, {Rezac}, {Ricci}, {Richards}, {Richardson}, {Richardson},
  {Riemenschneider}, {Riles}, {Rinaldi}, {Rink}, {Rizzo}, {Robertson}, {Robie},
  {Robinet}, {Rocchi}, {Rodriguez}, {Rolland}, {Rollins}, {Romanelli},
  {Romano}, {Romel}, {Romero-Rodr{\'\i}guez}, {Romero-Shaw}, {Romie},
  {Ronchini}, {Rosa}, {Rose}, {Rosi{\'n}ska}, {Ross}, {Rowan}, {Rowlinson},
  {Roy}, {Roy}, {Roy}, {Rozza}, {Ruggi}, {Ruiz-Rocha}, {Ryan}, {Sachdev},
  {Sadecki}, {Sadiq}, {Sago}, {Saito}, {Saito}, {Sakai}, {Sakai},
  {Sakellariadou}, {Sakuno}, {Salafia}, {Salconi}, {Saleem}, {Salemi},
  {Samajdar}, {Sanchez}, {Sanchez}, {Sanchez}, {Sanchis-Gual}, {Sanders},
  {Sanuy}, {Saravanan}, {Sarin}, {Sassolas}, {Satari}, {Sathyaprakash}, {Sato},
  {Sato}, {Sauter}, {Savage}, {Sawada}, {Sawant}, {Sawant}, {Sayah},
  {Schaetzl}, {Scheel}, {Scheuer}, {Schiworski}, {Schmidt}, {Schmidt},
  {Schnabel}, {Schneewind}, {Schofield}, {Sch{\"o}nbeck}, {Schulte}, {Schutz},
  {Schwartz}, {Scott}, {Scott}, {Seglar-Arroyo}, {Sekiguchi}, {Sekiguchi},
  {Sellers}, {Sengupta}, {Sentenac}, {Seo}, {Sequino}, {Sergeev}, {Setyawati},
  {Shaffer}, {Shahriar}, {Shams}, {Shao}, {Sharma}, {Sharma}, {Shawhan},
  {Shcheblanov}, {Shibagaki}, {Shikauchi}, {Shimizu}, {Shimoda}, {Shimode},
  {Shinkai}, {Shishido}, {Shoda}, {Shoemaker}, {Shoemaker}, {ShyamSundar},
  {Sieniawska}, {Sigg}, {Singer}, {Singh}, {Singh}, {Singha}, {Sintes},
  {Sipala}, {Skliris}, {Slagmolen}, {Slaven-Blair}, {Smetana}, {Smith},
  {Smith}, {Soldateschi}, {Somala}, {Somiya}, {Son}, {Soni}, {Soni}, {Sordini},
  {Sorrentino}, {Sorrentino}, {Sotani}, {Soulard}, {Souradeep}, {Sowell},
  {Spagnuolo}, {Spencer}, {Spera}, {Srinivasan}, {Srivastava}, {Srivastava},
  {Staats}, {Stachie}, {Steer}, {Steinhoff}, {Steinlechner}, {Steinlechner},
  {Stevenson}, {Stops}, {Stover}, {Strain}, {Strang}, {Stratta}, {Strunk},
  {Sturani}, {Stuver}, {Sudhagar}, {Sudhir}, {Sugimoto}, {Suh}, {Sullivan},
  {Sullivan}, {Summerscales}, {Sun}, {Sun}, {Sunil}, {Sur}, {Suresh}, {Sutton},
  {Suzuki}, {Suzuki}, {Swinkels}, {Szczepa{\'n}czyk}, {Szewczyk}, {Tacca},
  {Tagoshi}, {Tait}, {Takahashi}, {Takahashi}, {Takamori}, {Takano}, {Takeda},
  {Takeda}, {Talbot}, {Talbot}, {Tanaka}, {Tanaka}, {Tanaka}, {Tanaka},
  {Tanaka}, {Tanasijczuk}, {Tanioka}, {Tanner}, {Tao}, {Tao}, {Tapia San
  Mart{\'\i}n}, {Taranto}, {Tasson}, {Telada}, {Tenorio}, {Terhune},
  {Terkowski}, {Thirugnanasambandam}, {Thomas}, {Thomas}, {Thomas}, {Thompson},
  {Thondapu}, {Thorne}, {Thrane}, {Tiwari}, {Tiwari}, {Tiwari}, {Toivonen},
  {Toland}, {Tolley}, {Tomaru}, {Tomigami}, {Tomura}, {Tonelli},
  {Torres-Forn{\'e}}, {Torrie}, {Tosta e Melo}, {T{\"o}yr{\"a}}, {Trapananti},
  {Travasso}, {Traylor}, {Trevor}, {Tringali}, {Tripathee}, {Troiano},
  {Trovato}, {Trozzo}, {Trudeau}, {Tsai}, {Tsai}, {Tsang}, {Tsang}, {Tsao},
  {Tse}, {Tso}, {Tsubono}, {Tsuchida}, {Tsukada}, {Tsuna}, {Tsutsui},
  {Tsuzuki}, {Turbang}, {Turconi}, {Tuyenbayev}, {Ubhi}, {Uchikata},
  {Uchiyama}, {Udall}, {Ueda}, {Uehara}, {Ueno}, {Ueshima}, {Unnikrishnan},
  {Uraguchi}, {Urban}, {Ushiba}, {Utina}, {Vahlbruch}, {Vajente}, {Vajpeyi},
  {Valdes}, {Valentini}, {Valsan}, {van Bakel}, {van Beuzekom}, {van den
  Brand}, {Van Den Broeck}, {Vander-Hyde}, {van der Schaaf}, {van Heijningen},
  {Vanosky}, {van Putten}, {van Remortel}, {Vardaro}, {Vargas}, {Varma},
  {Vas{\'u}th}, {Vecchio}, {Vedovato}, {Veitch}, {Veitch}, {Venneberg},
  {Venugopalan}, {Verkindt}, {Verma}, {Verma}, {Veske}, {Vetrano},
  {Vicer{\'e}}, {Vidyant}, {Viets}, {Vijaykumar}, {Villa-Ortega}, {Vinet},
  {Virtuoso}, {Vitale}, {Vo}, {Vocca}, {von Reis}, {von Wrangel}, {Vorvick},
  {Vyatchanin}, {Wade}, {Wade}, {Wagner}, {Walet}, {Walker}, {Wallace},
  {Wallace}, {Walsh}, {Wang}, {Wang}, {Wang}, {Ward}, {Warner}, {Was},
  {Washimi}, {Washington}, {Watchi}, {Weaver}, {Webster}, {Weinert},
  {Weinstein}, {Weiss}, {Weller}, {Weller}, {Wellmann}, {Wen}, {We{\ss}els},
  {Wette}, {Whelan}, {White}, {Whiting}, {Whittle}, {Wilken}, {Williams},
  {Williams}, {Williams}, {Williamson}, {Willis}, {Willke}, {Wilson},
  {Winkler}, {Wipf}, {Wlodarczyk}, {Woan}, {Woehler}, {Wofford}, {Wong}, {Wu},
  {Wu}, {Wu}, {Wu}, {Wysocki}, {Xiao}, {Xu}, {Yamada}, {Yamamoto}, {Yamamoto},
  {Yamamoto}, {Yamamoto}, {Yamashita}, {Yamazaki}, {Yang}, {Yang}, {Yang},
  {Yang}, {Yang}, {Yap}, {Yeeles}, {Yelikar}, {Ying}, {Yokogawa}, {Yokoyama},
  {Yokozawa}, {Yoo}, {Yoshioka}, {Yu}, {Yu}, {Yuzurihara}, {Zadro{\.z}ny},
  {Zanolin}, {Zeidler}, {Zelenova}, {Zendri}, {Zevin}, {Zhan}, {Zhang},
  {Zhang}, {Zhang}, {Zhang}, {Zhang}, {Zhao}, {Zhao}, {Zhao}, {Zhao}, {Zheng},
  {Zhou}, {Zhou}, {Zhu}, {Zhu}, {Zimmerman}, {Zlochower}, {Zucker}, \&
  {Zweizig}}]{LSC2021}
{The LIGO Scientific Collaboration}, {the Virgo Collaboration}, {the KAGRA
  Collaboration}, {et~al.} 2021{\natexlab{a}}, arXiv e-prints,
  arXiv:2111.03606.
\newblock \doarXiv{2111.03606}

\bibitem[{{The LIGO Scientific Collaboration} {et~al.}(2021{\natexlab{b}}){The
  LIGO Scientific Collaboration}, {the Virgo Collaboration}, {the KAGRA
  Collaboration}, {Abbott}, {Abbott}, {Acernese}, {Ackley}, {Adams},
  {Adhikari}, {Adhikari}, {Adya}, {Affeldt}, {Agarwal}, {Agathos}, {Agatsuma},
  {Aggarwal}, {Aguiar}, {Aiello}, {Ain}, {Ajith}, {Akutsu}, {Albanesi},
  {Allocca}, {Altin}, {Amato}, {Anand}, {Anand}, {Ananyeva}, {Anderson},
  {Anderson}, {Ando}, {Andrade}, {Andres}, {Andri{\'c}}, {Angelova}, {Ansoldi},
  {Antelis}, {Antier}, {Antonini}, {Appert}, {Arai}, {Arai}, {Arai}, {Araki},
  {Araya}, {Araya}, {Areeda}, {Ar{\`e}ne}, {Aritomi}, {Arnaud}, {Aronson},
  {Arun}, {Asada}, {Asali}, {Ashton}, {Aso}, {Assiduo}, {Aston}, {Astone},
  {Aubin}, {Austin}, {Babak}, {Badaracco}, {Bader}, {Badger}, {Bae}, {Bae},
  {Baer}, {Bagnasco}, {Bai}, {Baiotti}, {Baird}, {Bajpai}, {Ball}, {Ballardin},
  {Ballmer}, {Balsamo}, {Baltus}, {Banagiri}, {Bankar}, {Barayoga}, {Barbieri},
  {Barish}, {Barker}, {Barneo}, {Barone}, {Barr}, {Barsotti}, {Barsuglia},
  {Barta}, {Bartlett}, {Barton}, {Bartos}, {Bassiri}, {Basti}, {Bawaj},
  {Bayley}, {Baylor}, {Bazzan}, {B{\'e}csy}, {Bedakihale}, {Bejger},
  {Belahcene}, {Benedetto}, {Beniwal}, {Bennett}, {Bentley}, {BenYaala},
  {Bergamin}, {Berger}, {Bernuzzi}, {Berry}, {Bersanetti}, {Bertolini},
  {Betzwieser}, {Beveridge}, {Bhandare}, {Bhardwaj}, {Bhattacharjee},
  {Bhaumik}, {Bilenko}, {Billingsley}, {Bini}, {Birney}, {Birnholtz},
  {Biscans}, {Bischi}, {Biscoveanu}, {Bisht}, {Biswas}, {Bitossi}, {Bizouard},
  {Blackburn}, {Blair}, {Blair}, {Blair}, {Bobba}, {Bode}, {Boer}, {Bogaert},
  {Boldrini}, {Bonavena}, {Bondu}, {Bonilla}, {Bonnand}, {Booker}, {Boom},
  {Bork}, {Boschi}, {Bose}, {Bose}, {Bossilkov}, {Boudart}, {Bouffanais},
  {Bozzi}, {Bradaschia}, {Brady}, {Bramley}, {Branch}, {Branchesi}, {Brau},
  {Breschi}, {Briant}, {Briggs}, {Brillet}, {Brinkmann}, {Brockill}, {Brooks},
  {Brooks}, {Brown}, {Brunett}, {Bruno}, {Bruntz}, {Bryant}, {Bulik}, {Bulten},
  {Buonanno}, {Buscicchio}, {Buskulic}, {Buy}, {Byer}, {Cadonati}, {Cagnoli},
  {Cahillane}, {Calder{\'o}n Bustillo}, {Callaghan}, {Callister}, {Calloni},
  {Cameron}, {Camp}, {Canepa}, {Canevarolo}, {Cannavacciuolo}, {Cannon}, {Cao},
  {Cao}, {Capocasa}, {Capote}, {Carapella}, {Carbognani}, {Carlin}, {Carney},
  {Carpinelli}, {Carrillo}, {Carullo}, {Carver}, {Casanueva Diaz}, {Casentini},
  {Castaldi}, {Caudill}, {Cavagli{\`a}}, {Cavalier}, {Cavalieri}, {Ceasar},
  {Cella}, {Cerd{\'a}-Dur{\'a}n}, {Cesarini}, {Chaibi}, {Chakravarti},
  {Chalathadka Subrahmanya}, {Champion}, {Chan}, {Chan}, {Chan}, {Chan},
  {Chan}, {Chandra}, {Chanial}, {Chao}, {Charlton}, {Chase},
  {Chassande-Mottin}, {Chatterjee}, {Chatterjee}, {Chatterjee}, {Chaturvedi},
  {Chaty}, {Chatziioannou}, {Chen}, {Chen}, {Chen}, {Chen}, {Chen}, {Chen},
  {Chen}, {Chen}, {Cheng}, {Cheong}, {Cheung}, {Chia}, {Chiadini}, {Chiang},
  {Chiarini}, {Chierici}, {Chincarini}, {Chiofalo}, {Chiummo}, {Cho}, {Cho},
  {Choudhary}, {Choudhary}, {Christensen}, {Chu}, {Chu}, {Chu}, {Chua},
  {Chung}, {Ciani}, {Ciecielag}, {Cie{\'s}lar}, {Cifaldi}, {Ciobanu}, {Ciolfi},
  {Cipriano}, {Cirone}, {Clara}, {Clark}, {Clark}, {Clarke}, {Clearwater},
  {Clesse}, {Cleva}, {Coccia}, {Codazzo}, {Cohadon}, {Cohen}, {Cohen},
  {Colleoni}, {Collette}, {Colombo}, {Colpi}, {Compton}, {Constancio}, {Conti},
  {Cooper}, {Corban}, {Corbitt}, {Cordero-Carri{\'o}n}, {Corezzi}, {Corley},
  {Cornish}, {Corre}, {Corsi}, {Cortese}, {Costa}, {Cotesta}, {Coughlin},
  {Coulon}, {Countryman}, {Cousins}, {Couvares}, {Coward}, {Cowart}, {Coyne},
  {Coyne}, {Creighton}, {Creighton}, {Criswell}, {Croquette}, {Crowder},
  {Cudell}, {Cullen}, {Cumming}, {Cummings}, {Cunningham}, {Cuoco},
  {Cury{\l}o}, {Dabadie}, {Dal Canton}, {Dall'Osso}, {D{\'a}lya}, {Dana},
  {DaneshgaranBajastani}, {D'Angelo}, {Danilishin}, {D'Antonio}, {Danzmann},
  {Darsow-Fromm}, {Dasgupta}, {Datrier}, {Datta}, {Dattilo}, {Dave}, {Davier},
  {Davies}, {Davis}, {Davis}, {Daw}, {Dean}, {DeBra}, {Deenadayalan},
  {Degallaix}, {De Laurentis}, {Del{\'e}glise}, {Del Favero}, {De Lillo}, {De
  Lillo}, {Del Pozzo}, {DeMarchi}, {De Matteis}, {D'Emilio}, {Demos}, {Dent},
  {Depasse}, {De Pietri}, {De Rosa}, {De Rossi}, {DeSalvo}, {De Simone},
  {Dhurandhar}, {D{\'\i}az}, {Diaz-Ortiz}, {Didio}, {Dietrich}, {Di Fiore}, {Di
  Fronzo}, {Di Giorgio}, {Di Giovanni}, {Di Giovanni}, {Di Girolamo}, {Di
  Lieto}, {Ding}, {Di Pace}, {Di Palma}, {Di Renzo}, {Divakarla}, {Dmitriev},
  {Doctor}, {D'Onofrio}, {Donovan}, {Dooley}, {Doravari}, {Dorrington},
  {Drago}, {Driggers}, {Drori}, {Ducoin}, {Dupej}, {Durante}, {D'Urso},
  {Duverne}, {Dwyer}, {Eassa}, {Easter}, {Ebersold}, {Eckhardt}, {Eddolls},
  {Edelman}, {Edo}, {Edy}, {Effler}, {Eguchi}, {Eichholz}, {Eikenberry},
  {Eisenmann}, {Eisenstein}, {Ejlli}, {Engelby}, {Enomoto}, {Errico}, {Essick},
  {Estell{\'e}s}, {Estevez}, {Etienne}, {Etzel}, {Evans}, {Evans}, {Ewing},
  {Fafone}, {Fair}, {Fairhurst}, {Farah}, {Farinon}, {Farr}, {Farr}, {Farrow},
  {Fauchon-Jones}, {Favaro}, {Favata}, {Fays}, {Fazio}, {Feicht}, {Fejer},
  {Fenyvesi}, {Ferguson}, {Fernandez-Galiana}, {Ferrante}, {Ferreira},
  {Fidecaro}, {Figura}, {Fiori}, {Fishbach}, {Fisher}, {Fittipaldi}, {Fiumara},
  {Flaminio}, {Floden}, {Fong}, {Font}, {Fornal}, {Forsyth}, {Franke},
  {Frasca}, {Frasconi}, {Frederick}, {Freed}, {Frei}, {Freise}, {Frey},
  {Fritschel}, {Frolov}, {Fronz{\'e}}, {Fujii}, {Fujikawa}, {Fukunaga},
  {Fukushima}, {Fulda}, {Fyffe}, {Gabbard}, {Gadre}, {Gair}, {Gais},
  {Galaudage}, {Gamba}, {Ganapathy}, {Ganguly}, {Gao}, {Gaonkar}, {Garaventa},
  {Garc{\'\i}a-N{\'u}{\~n}ez}, {Garc{\'\i}a-Quir{\'o}s}, {Garufi}, {Gateley},
  {Gaudio}, {Gayathri}, {Ge}, {Gemme}, {Gennai}, {George}, {Gerberding},
  {Gergely}, {Gewecke}, {Ghonge}, {Ghosh}, {Ghosh}, {Ghosh}, {Ghosh},
  {Giacomazzo}, {Giacoppo}, {Giaime}, {Giardina}, {Gibson}, {Gier}, {Giesler},
  {Giri}, {Gissi}, {Glanzer}, {Gleckl}, {Godwin}, {Goetz}, {Goetz}, {Gohlke},
  {Golomb}, {Goncharov}, {Gonz{\'a}lez}, {Gopakumar}, {Gosselin}, {Gouaty},
  {Gould}, {Grace}, {Grado}, {Granata}, {Granata}, {Grant}, {Gras}, {Grassia},
  {Gray}, {Gray}, {Greco}, {Green}, {Green}, {Gretarsson}, {Gretarsson},
  {Griffith}, {Griffiths}, {Griggs}, {Grignani}, {Grimaldi}, {Grimm}, {Grote},
  {Grunewald}, {Gruning}, {Guerra}, {Guidi}, {Guimaraes}, {Guix{\'e}},
  {Gulati}, {Guo}, {Guo}, {Gupta}, {Gupta}, {Gupta}, {Gustafson}, {Gustafson},
  {Guzman}, {Ha}, {Haegel}, {Hagiwara}, {Haino}, {Halim}, {Hall}, {Hamilton},
  {Hammond}, {Han}, {Haney}, {Hanks}, {Hanna}, {Hannam}, {Hannuksela},
  {Hansen}, {Hansen}, {Hanson}, {Harder}, {Hardwick}, {Haris}, {Harms},
  {Harry}, {Harry}, {Hartwig}, {Hasegawa}, {Haskell}, {Hasskew}, {Haster},
  {Hattori}, {Haughian}, {Hayakawa}, {Hayama}, {Hayes}, {Healy}, {Heidmann},
  {Heidt}, {Heintze}, {Heinze}, {Heinzel}, {Heitmann}, {Hellman}, {Hello},
  {Helmling-Cornell}, {Hemming}, {Hendry}, {Heng}, {Hennes}, {Hennig},
  {Hennig}, {Hernandez}, {Hernandez Vivanco}, {Heurs}, {Hild}, {Hill},
  {Himemoto}, {Hines}, {Hiranuma}, {Hirata}, {Hirose}, {Hochheim}, {Hofman},
  {Hohmann}, {Holcomb}, {Holland}, {Hollows}, {Holmes}, {Holt}, {Holz}, {Hong},
  {Hopkins}, {Hough}, {Hourihane}, {Howell}, {Hoy}, {Hoyland}, {Hreibi},
  {Hsieh}, {Hsu}, {Huang}, {Huang}, {Huang}, {Huang}, {Huang}, {Huang},
  {H{\"u}bner}, {Huddart}, {Hughey}, {Hui}, {Hui}, {Husa}, {Huttner},
  {Huxford}, {Huynh-Dinh}, {Ide}, {Idzkowski}, {Iess}, {Ikenoue}, {Imam},
  {Inayoshi}, {Ingram}, {Inoue}, {Ioka}, {Isi}, {Isleif}, {Ito}, {Itoh},
  {Iyer}, {Izumi}, {JaberianHamedan}, {Jacqmin}, {Jadhav}, {Jadhav}, {James},
  {Jan}, {Jani}, {Janquart}, {Janssens}, {Janthalur}, {Jaranowski}, {Jariwala},
  {Jaume}, {Jenkins}, {Jenner}, {Jeon}, {Jeunon}, {Jia}, {Jin}, {Johns},
  {Jones}, {Jones}, {Jones}, {Jones}, {Jones}, {Jonker}, {Ju}, {Jung}, {Jung},
  {Junker}, {Juste}, {Kaihotsu}, {Kajita}, {Kakizaki}, {Kalaghatgi},
  {Kalogera}, {Kamai}, {Kamiizumi}, {Kanda}, {Kandhasamy}, {Kang}, {Kanner},
  {Kao}, {Kapadia}, {Kapasi}, {Karat}, {Karathanasis}, {Karki}, {Kashyap},
  {Kasprzack}, {Kastaun}, {Katsanevas}, {Katsavounidis}, {Katzman}, {Kaur},
  {Kawabe}, {Kawaguchi}, {Kawai}, {Kawasaki}, {K{\'e}f{\'e}lian}, {Keitel},
  {Key}, {Khadka}, {Khalili}, {Khan}, {Khazanov}, {Khetan}, {Khursheed},
  {Kijbunchoo}, {Kim}, {Kim}, {Kim}, {Kim}, {Kim}, {Kim}, {Kimball}, {Kimura},
  {Kinley-Hanlon}, {Kirchhoff}, {Kissel}, {Kita}, {Kitazawa}, {Kleybolte},
  {Klimenko}, {Knee}, {Knowles}, {Knyazev}, {Koch}, {Koekoek}, {Kojima},
  {Kokeyama}, {Koley}, {Kolitsidou}, {Kolstein}, {Komori}, {Kondrashov},
  {Kong}, {Kontos}, {Koper}, {Korobko}, {Kotake}, {Kovalam}, {Kozak},
  {Kozakai}, {Kozu}, {Kringel}, {Krishnendu}, {Kr{\'o}lak}, {Kuehn}, {Kuei},
  {Kuijer}, {Kumar}, {Kumar}, {Kumar}, {Kumar}, {Kume}, {Kuns}, {Kuo}, {Kuo},
  {Kuromiya}, {Kuroyanagi}, {Kusayanagi}, {Kuwahara}, {Kwak}, {Lagabbe},
  {Laghi}, {Lalande}, {Lam}, {Lamberts}, {Landry}, {Landry}, {Lane}, {Lang},
  {Lange}, {Lantz}, {La Rosa}, {Lartaux-Vollard}, {Lasky}, {Laxen},
  {Lazzarini}, {Lazzaro}, {Leaci}, {Leavey}, {Lecoeuche}, {Lee}, {Lee}, {Lee},
  {Lee}, {Lee}, {Lee}, {Lehmann}, {Lema{\^\i}tre}, {Leonardi}, {Leroy},
  {Letendre}, {Levesque}, {Levin}, {Leviton}, {Leyde}, {Li}, {Li}, {Li}, {Li},
  {Li}, {Li}, {Lin}, {Lin}, {Lin}, {Lin}, {Lin}, {Linde}, {Linker}, {Linley},
  {Littenberg}, {Liu}, {Liu}, {Liu}, {Liu}, {Llamas}, {Llorens-Monteagudo},
  {Lo}, {Lockwood}, {London}, {Longo}, {Lopez}, {Lopez Portilla}, {Lorenzini},
  {Loriette}, {Lormand}, {Losurdo}, {Lott}, {Lough}, {Lousto}, {Lovelace},
  {Lucaccioni}, {L{\"u}ck}, {Lumaca}, {Lundgren}, {Luo}, {Lynam}, {Macas},
  {MacInnis}, {Macleod}, {MacMillan}, {Macquet}, {Maga{\~n}a Hernandez},
  {Magazz{\`u}}, {Magee}, {Maggiore}, {Magnozzi}, {Mahesh}, {Majorana},
  {Makarem}, {Maksimovic}, {Maliakal}, {Malik}, {Man}, {Mandic}, {Mangano},
  {Mango}, {Mansell}, {Manske}, {Mantovani}, {Mapelli}, {Marchesoni},
  {Marchio}, {Marion}, {Mark}, {M{\'a}rka}, {M{\'a}rka}, {Markakis},
  {Markosyan}, {Markowitz}, {Maros}, {Marquina}, {Marsat}, {Martelli},
  {Martin}, {Martin}, {Martinez}, {Martinez}, {Martinez}, {Martinovic},
  {Martynov}, {Marx}, {Masalehdan}, {Mason}, {Massera}, {Masserot},
  {Massinger}, {Masso-Reid}, {Mastrogiovanni}, {Matas}, {Mateu-Lucena},
  {Matichard}, {Matiushechkina}, {Mavalvala}, {McCann}, {McCarthy},
  {McClelland}, {McClincy}, {McCormick}, {McCuller}, {McGhee}, {McGuire},
  {McIsaac}, {McIver}, {McRae}, {McWilliams}, {Meacher}, {Mehmet}, {Mehta},
  {Meijer}, {Melatos}, {Melchor}, {Mendell}, {Menendez-Vazquez}, {Menoni},
  {Mercer}, {Mereni}, {Merfeld}, {Merilh}, {Merritt}, {Merzougui}, {Meshkov},
  {Messenger}, {Messick}, {Meyers}, {Meylahn}, {Mhaske}, {Miani}, {Miao},
  {Michaloliakos}, {Michel}, {Michimura}, {Middleton}, {Milano}, {Miller},
  {Miller}, {Miller}, {Miller}, {Millhouse}, {Mills}, {Milotti}, {Minazzoli},
  {Minenkov}, {Mio}, {Mir}, {Miravet-Ten{\'e}s}, {Mishra}, {Mishra}, {Mistry},
  {Mitra}, {Mitrofanov}, {Mitselmakher}, {Mittleman}, {Miyakawa}, {Miyamoto},
  {Miyazaki}, {Miyo}, {Miyoki}, {Mo}, {Moguel}, {Mogushi}, {Mohapatra},
  {Mohite}, {Molina}, {Molina-Ruiz}, {Mondin}, {Montani}, {Moore}, {Moraru},
  {Morawski}, {More}, {Moreno}, {Moreno}, {Mori}, {Morisaki}, {Moriwaki},
  {Mours}, {Mow-Lowry}, {Mozzon}, {Muciaccia}, {Mukherjee}, {Mukherjee},
  {Mukherjee}, {Mukherjee}, {Mukherjee}, {Mukund}, {Mullavey}, {Munch},
  {Mu{\~n}iz}, {Murray}, {Musenich}, {Muusse}, {Nadji}, {Nagano}, {Nagano},
  {Nagar}, {Nakamura}, {Nakano}, {Nakano}, {Nakashima}, {Nakayama}, {Napolano},
  {Nardecchia}, {Narikawa}, {Naticchioni}, {Nayak}, {Nayak}, {Negishi}, {Neil},
  {Neilson}, {Nelemans}, {Nelson}, {Nery}, {Neubauer}, {Neunzert}, {Ng}, {Ng},
  {Nguyen}, {Nguyen}, {Nguyen}, {Nguyen Quynh}, {Ni}, {Nichols}, {Nishizawa},
  {Nissanke}, {Nitoglia}, {Nocera}, {Norman}, {North}, {Nozaki}, {Nuttall},
  {Oberling}, {O'Brien}, {Obuchi}, {O'Dell}, {Oelker}, {Ogaki}, {Oganesyan},
  {Oh}, {Oh}, {Oh}, {Ohashi}, {Ohishi}, {Ohkawa}, {Ohme}, {Ohta}, {Okada},
  {Okutani}, {Okutomi}, {Olivetto}, {Oohara}, {Ooi}, {Oram}, {O'Reilly},
  {Ormiston}, {Ormsby}, {Ortega}, {O'Shaughnessy}, {O'Shea}, {Oshino},
  {Ossokine}, {Osthelder}, {Otabe}, {Ottaway}, {Overmier}, {Pace}, {Pagano},
  {Page}, {Pagliaroli}, {Pai}, {Pai}, {Palamos}, {Palashov}, {Palomba}, {Pan},
  {Pan}, {Panda}, {Pang}, {Pang}, {Pankow}, {Pannarale}, {Pant}, {Panther},
  {Paoletti}, {Paoli}, {Paolone}, {Parisi}, {Park}, {Park}, {Parker},
  {Pascucci}, {Pasqualetti}, {Passaquieti}, {Passuello}, {Patel}, {Pathak},
  {Patricelli}, {Patron}, {Paul}, {Payne}, {Pedraza}, {Pegoraro}, {Pele},
  {Pe{\~n}a Arellano}, {Penn}, {Perego}, {Pereira}, {Pereira}, {Perez},
  {P{\'e}rigois}, {Perkins}, {Perreca}, {Perri{\`e}s}, {Petermann},
  {Petterson}, {Pfeiffer}, {Pham}, {Phukon}, {Piccinni}, {Pichot},
  {Piendibene}, {Piergiovanni}, {Pierini}, {Pierro}, {Pillant}, {Pillas},
  {Pilo}, {Pinard}, {Pinto}, {Pinto}, {Piotrzkowski}, {Pirello}, {Pitkin},
  {Placidi}, {Planas}, {Plastino}, {Pluchar}, {Poggiani}, {Polini}, {Pong},
  {Ponrathnam}, {Popolizio}, {Porter}, {Poulton}, {Powell}, {Pracchia},
  {Pradier}, {Prajapati}, {Prasai}, {Prasanna}, {Pratten}, {Principe}, {Prodi},
  {Prokhorov}, {Prosposito}, {Prudenzi}, {Puecher}, {Punturo}, {Puosi},
  {Puppo}, {P{\"u}rrer}, {Qi}, {Quetschke}, {Quitzow-James}, {Raab},
  {Raaijmakers}, {Radkins}, {Radulesco}, {Raffai}, {Rail}, {Raja}, {Rajan},
  {Ramirez}, {Ramirez}, {Ramos-Buades}, {Rana}, {Rapagnani}, {Rapol}, {Ray},
  {Raymond}, {Raza}, {Razzano}, {Read}, {Rees}, {Regimbau}, {Rei}, {Reid},
  {Reid}, {Reitze}, {Relton}, {Renzini}, {Rettegno}, {Rezac}, {Ricci},
  {Richards}, {Richardson}, {Richardson}, {Riemenschneider}, {Riles},
  {Rinaldi}, {Rink}, {Rizzo}, {Robertson}, {Robie}, {Robinet}, {Rocchi},
  {Rodriguez}, {Rolland}, {Rollins}, {Romanelli}, {Romano}, {Romel},
  {Romero-Rodr{\'\i}guez}, {Romero-Shaw}, {Romie}, {Ronchini}, {Rosa}, {Rose},
  {Rosi{\'n}ska}, {Ross}, {Rowan}, {Rowlinson}, {Roy}, {Roy}, {Roy}, {Rozza},
  {Ruggi}, {Ryan}, {Sachdev}, {Sadecki}, {Sadiq}, {Sago}, {Saito}, {Saito},
  {Sakai}, {Sakai}, {Sakellariadou}, {Sakuno}, {Salafia}, {Salconi}, {Saleem},
  {Salemi}, {Samajdar}, {Sanchez}, {Sanchez}, {Sanchez}, {Sanchis-Gual},
  {Sanders}, {Sanuy}, {Saravanan}, {Sarin}, {Sassolas}, {Satari},
  {Sathyaprakash}, {Sato}, {Sato}, {Sauter}, {Savage}, {Sawada}, {Sawant},
  {Sawant}, {Sayah}, {Schaetzl}, {Scheel}, {Scheuer}, {Schiworski}, {Schmidt},
  {Schmidt}, {Schnabel}, {Schneewind}, {Schofield}, {Sch{\"o}nbeck}, {Schulte},
  {Schutz}, {Schwartz}, {Scott}, {Scott}, {Seglar-Arroyo}, {Sekiguchi},
  {Sekiguchi}, {Sellers}, {Sengupta}, {Sentenac}, {Seo}, {Sequino}, {Sergeev},
  {Setyawati}, {Shaffer}, {Shahriar}, {Shams}, {Shao}, {Sharma}, {Sharma},
  {Shawhan}, {Shcheblanov}, {Shibagaki}, {Shikauchi}, {Shimizu}, {Shimoda},
  {Shimode}, {Shinkai}, {Shishido}, {Shoda}, {Shoemaker}, {Shoemaker},
  {ShyamSundar}, {Sieniawska}, {Sigg}, {Singer}, {Singh}, {Singh}, {Singha},
  {Sintes}, {Sipala}, {Skliris}, {Slagmolen}, {Slaven-Blair}, {Smetana},
  {Smith}, {Smith}, {Soldateschi}, {Somala}, {Somiya}, {Son}, {Soni}, {Soni},
  {Sordini}, {Sorrentino}, {Sorrentino}, {Sotani}, {Soulard}, {Souradeep},
  {Sowell}, {Spagnuolo}, {Spencer}, {Spera}, {Srinivasan}, {Srivastava},
  {Srivastava}, {Staats}, {Stachie}, {Steer}, {Steinlechner}, {Steinlechner},
  {Stops}, {Stover}, {Strain}, {Strang}, {Stratta}, {Strunk}, {Sturani},
  {Stuver}, {Sudhagar}, {Sudhir}, {Sugimoto}, {Suh}, {Summerscales}, {Sun},
  {Sun}, {Sunil}, {Sur}, {Suresh}, {Sutton}, {Suzuki}, {Suzuki}, {Swinkels},
  {Szczepa{\'n}czyk}, {Szewczyk}, {Tacca}, {Tagoshi}, {Tait}, {Takahashi},
  {Takahashi}, {Takamori}, {Takano}, {Takeda}, {Takeda}, {Talbot}, {Talbot},
  {Tanaka}, {Tanaka}, {Tanaka}, {Tanaka}, {Tanaka}, {Tanasijczuk}, {Tanioka},
  {Tanner}, {Tao}, {Tao}, {Tapia San Mart{\'\i}n}, {Taranto}, {Tasson},
  {Telada}, {Tenorio}, {Terhune}, {Terkowski}, {Thirugnanasambandam}, {Thomas},
  {Thomas}, {Thompson}, {Thondapu}, {Thorne}, {Thrane}, {Tiwari}, {Tiwari},
  {Tiwari}, {Toivonen}, {Toland}, {Tolley}, {Tomaru}, {Tomigami}, {Tomura},
  {Tonelli}, {Torres-Forn{\'e}}, {Torrie}, {Tosta e Melo}, {T{\"o}yr{\"a}},
  {Trapananti}, {Travasso}, {Traylor}, {Trevor}, {Tringali}, {Tripathee},
  {Troiano}, {Trovato}, {Trozzo}, {Trudeau}, {Tsai}, {Tsai}, {Tsang}, {Tsang},
  {Tsao}, {Tse}, {Tso}, {Tsubono}, {Tsuchida}, {Tsukada}, {Tsuna}, {Tsutsui},
  {Tsuzuki}, {Turbang}, {Turconi}, {Tuyenbayev}, {Ubhi}, {Uchikata},
  {Uchiyama}, {Udall}, {Ueda}, {Uehara}, {Ueno}, {Ueshima}, {Unnikrishnan},
  {Uraguchi}, {Urban}, {Ushiba}, {Utina}, {Vahlbruch}, {Vajente}, {Vajpeyi},
  {Valdes}, {Valentini}, {Valsan}, {van Bakel}, {van Beuzekom}, {van den
  Brand}, {Van Den Broeck}, {Vander-Hyde}, {van der Schaaf}, {van Heijningen},
  {Vanosky}, {van Putten}, {van Remortel}, {Vardaro}, {Vargas}, {Varma},
  {Vas{\'u}th}, {Vecchio}, {Vedovato}, {Veitch}, {Veitch}, {Venneberg},
  {Venugopalan}, {Verkindt}, {Verma}, {Verma}, {Veske}, {Vetrano},
  {Vicer{\'e}}, {Vidyant}, {Viets}, {Vijaykumar}, {Villa-Ortega}, {Vinet},
  {Virtuoso}, {Vitale}, {Vo}, {Vocca}, {von Reis}, {von Wrangel}, {Vorvick},
  {Vyatchanin}, {Wade}, {Wade}, {Wagner}, {Walet}, {Walker}, {Wallace},
  {Wallace}, {Walsh}, {Wang}, {Wang}, {Wang}, {Ward}, {Warner}, {Was},
  {Washimi}, {Washington}, {Watchi}, {Weaver}, {Webster}, {Weinert},
  {Weinstein}, {Weiss}, {Weller}, {Wellmann}, {Wen}, {We{\ss}els}, {Wette},
  {Whelan}, {White}, {Whiting}, {Whittle}, {Wilken}, {Williams}, {Williams},
  {Williamson}, {Willis}, {Willke}, {Wilson}, {Winkler}, {Wipf}, {Wlodarczyk},
  {Woan}, {Woehler}, {Wofford}, {Wong}, {Wu}, {Wu}, {Wu}, {Wu}, {Wysocki},
  {Xiao}, {Xu}, {Yamada}, {Yamamoto}, {Yamamoto}, {Yamamoto}, {Yamamoto},
  {Yamashita}, {Yamazaki}, {Yang}, {Yang}, {Yang}, {Yang}, {Yang}, {Yap},
  {Yeeles}, {Yelikar}, {Ying}, {Yokogawa}, {Yokoyama}, {Yokozawa}, {Yoo},
  {Yoshioka}, {Yu}, {Yu}, {Yuzurihara}, {Zadro{\.z}ny}, {Zanolin}, {Zeidler},
  {Zelenova}, {Zendri}, {Zevin}, {Zhan}, {Zhang}, {Zhang}, {Zhang}, {Zhang},
  {Zhang}, {Zhao}, {Zhao}, {Zhao}, {Zhao}, {Zhou}, {Zhou}, {Zhu}, {Zhu},
  {Zimmerman}, {Zlochower}, {Zucker}, \& {Zweizig}}]{LSC2021a}
---. 2021{\natexlab{b}}, arXiv e-prints, arXiv:2111.03634.
\newblock \doarXiv{2111.03634}

\bibitem[{{Vigna-G{\'o}mez} {et~al.}(2021){Vigna-G{\'o}mez}, {Toonen},
  {Ramirez-Ruiz}, {Leigh}, {Riley}, \& {Haster}}]{VignaGomez2021}
{Vigna-G{\'o}mez}, A., {Toonen}, S., {Ramirez-Ruiz}, E., {et~al.} 2021, \apjl,
  907, L19, \dodoi{10.3847/2041-8213/abd5b7}

\bibitem[{{Wang} {et~al.}(2020){Wang}, {Stephan}, {Naoz}, {Hoang}, \&
  {Breivik}}]{Wang2020}
{Wang}, H., {Stephan}, A.~P., {Naoz}, S., {Hoang}, B.-M., \& {Breivik}, K.
  2020, arXiv e-prints, arXiv:2010.15841.
\newblock \doarXiv{2010.15841}

\bibitem[{{Ye} {et~al.}(2020){Ye}, {Fong}, {Kremer}, {Rodriguez}, {Chatterjee},
  {Fragione}, \& {Rasio}}]{Ye2020}
{Ye}, C.~S., {Fong}, W.-f., {Kremer}, K., {et~al.} 2020, \apjl, 888, L10,
  \dodoi{10.3847/2041-8213/ab5dc5}

\bibitem[{{Zhu} {et~al.}(2010){Zhu}, {Blanton}, \& {Moustakas}}]{Zhu2010}
{Zhu}, G., {Blanton}, M.~R., \& {Moustakas}, J. 2010, \apj, 722, 491,
  \dodoi{10.1088/0004-637X/722/1/491}

\end{thebibliography}



\end{document}